\documentclass[preprint2]{aastex61}

\begin{document}
\title{The largest M dwarfs flares from ASAS-SN}
\shorttitle{ASAS-SN M dwarf flares}

\author[0000-0002-7224-7702]{Sarah J. Schmidt}
\affil{Leibniz-Institute for Astrophysics Potsdam (AIP), An der Sternwarte 16, 14482, Potsdam, Germany}
\email{sjschmidt@aip.de}
\author{Benjamin J. Shappee}
\affil{Institute for Astronomy, University of Hawai'i, 2680 Woodlawn Drive, Honolulu, HI 96822, USA}
\affil{Hubble Fellow}
\affil{Carnegie-Princeton Fellow}
\affiliation{The Observatories of the Carnegie Institution for Science, 813 Santa Barbara St., Pasadena, CA 91101, USA}
\author{Jennifer L. van Saders}
\affil{Institute for Astronomy, University of Hawai'i, 2680 Woodlawn Drive, Honolulu, HI 96822, USA}
\affil{Carnegie-Princeton Fellow}
\affiliation{The Observatories of the Carnegie Institution for Science, 813 Santa Barbara St., Pasadena, CA 91101, USA}
\author{K.~Z. Stanek}
\affil{Department of Astronomy, The Ohio State University, 140 West 18th Avenue, Columbus, OH 43210, USA}
\affil{Center for Cosmology and Astroparticle Physics, The Ohio State University, 191 W. Woodruff Avenue, Columbus, OH 43210}
\author{Jonathan S. Brown}
\affil{Department of Astronomy, The Ohio State University, 140 West 18th Avenue, Columbus, OH 43210, USA}
\author{C.~S. Kochanek}
\affil{Department of Astronomy, The Ohio State University, 140 West 18th Avenue, Columbus, OH 43210, USA}
\affil{Center for Cosmology and Astroparticle Physics, The Ohio State University, 191 W. Woodruff Avenue, Columbus, OH 43210}
\author{Subo Dong} 
\affil{Kavli Institute for Astronomy and Astrophysics, Peking University, Yi He Yuan Road 5, Hai Dian District, Beijing 100871, China}
\author{Maria R. Drout}
\affil{Hubble Fellow}
\affil{Carnegie-Dunlap Fellow}
\affiliation{The Observatories of the Carnegie Institution for Science, 813 Santa Barbara St., Pasadena, CA 91101, USA}
\author{Stephan Frank}
\affil{Department of Astronomy, The Ohio State University, 140 West 18th Avenue, Columbus, OH 43210, USA}
\author[0000-0001-9206-3460]{T.~W.-S.~Holoien}
\affil{Carnegie Fellow}
\affiliation{The Observatories of the Carnegie Institution for Science, 813 Santa Barbara St., Pasadena, CA 91101, USA}
\author{Sean Johnson}
\affil{Hubble Fellow} 
\affil{Carnegie-Princeton Fellow}
\affil{Department of Astrophysics, Princeton University, Princeton, NJ, USA}
\author{Barry F. Madore} 
\affiliation{The Observatories of the Carnegie Institution for Science, 813 Santa Barbara St., Pasadena, CA 91101, USA}
\author{J. L. Prieto}
\affil{Nu\'cleo de Astronom\'ia de la Facultad de Ingenier\'ia y Ciencias, Universidad Diego Portales, Av. Ej\'ercito 441, Santiago, Chile}
\affil{Millennium Institute of Astrophysics, Santiago, Chile}
\author{Mark Seibert} 
\affiliation{The Observatories of the Carnegie Institution for Science, 813 Santa Barbara St., Pasadena, CA 91101, USA}
\author{Marja K. Seidel}
\affiliation{The Observatories of the Carnegie Institution for Science, 813 Santa Barbara St., Pasadena, CA 91101, USA}
\affil{Caltech-IPAC, Spitzer Science Center, 1200 E. California Blvd., Pasadena, CA 91125 USA}
\author{Gregory V. A. Simonian}
\affil{Department of Astronomy, The Ohio State University, 140 West 18th Avenue, Columbus, OH 43210, USA}

\begin{abstract} 
The All-Sky Automated Survey for Supernovae (ASAS-SN) is the only project in existence to scan the entire sky in optical light every $\sim$day, reaching a depth of $g\sim18$~mag. Over the course of its first four years of transient alerts (2013-2016), ASAS-SN observed 53 events classified as likely M dwarf flares. We present follow-up photometry and spectroscopy of all 53 candidates, confirming flare events on 47 M dwarfs, one K dwarf, and one L dwarf. The remaining four objects include a previously identified TT Tauri star, a young star with outbursts, and two objects too faint to confirm. A detailed examination of the 49 flare star light curves revealed an additional six flares on five stars, resulting in a total of 55 flares on 49 objects ranging in $V$-band contrast from $\Delta V = -1$ to $-10.2$~mags. Using an empirical flare model to estimate the unobserved portions of the flare light curve, we obtain lower limits on the $V$-band energy emitted during each flare, spanning $\log(E_V/{\rm ergs})=32$ to $35$, which are among the most energetic flares detected on M dwarfs. The ASAS-SN M-dwarf flare stars show a higher fraction of H$\alpha$ emission as well as stronger H$\alpha$ emission compared to M dwarfs selected without reference to activity, consistent with belonging to a population of more magnetically active stars. We also examined the distribution of tangential velocities, finding that the ASAS-SN flaring M dwarfs are likely to be members of the thin disk and are neither particularly young nor old.

\end{abstract}


\section{Introduction}
\label{sec:intro}
Stellar flares are dramatic multi-wavelength outbursts that originate in the reconnection of magnetic field lines at the surfaces of stars across the main sequence. A typical flare event stretches from the stellar photosphere to the corona, with high energy particle acceleration dominating radio and X-ray flare emission \citep[e.g.,][]{Lin1976,Osten2005}, while dense, hot material in the pre-flare photosphere and chromosphere emits at UV and optical wavelengths through a combination of thermal blackbody and atomic emission \citep[e.g.,][]{Hawley1991,Kowalski2013}. Most flares are rapid transients, characterized by an initial impulsive rise and decay (lasting a few to tens of minutes) followed by a more gradual exponential decay \citep[typically lasting hours; see e.g.,][]{Davenport2014a}. Flares are often observed on M dwarfs, both because these low-mass stars remain magnetically active and flaring even at ages of a few Gyr \citep{West2008,Hilton2010,Pineda2013}, and because the contrast of the hot ($\sim$10000~K), blue thermal component of the flare emission with the cool ($\sim$3000~K), red photosphere allows for observations of modest ($E\sim 10^{26}$~ergs) flares that require high precision photometry to detect on solar-type stars other than the Sun. So far, M dwarf flares have been observed to span a wide range of energies, from $E\sim 10^{26}$~ergs up to $10^{35}$--$10^{36}$~ergs \citep{Lacy1976,Kowalski2010,Davenport2016}. 

When a single star is observed long enough that a significant number of flares are detected (tens to hundreds), the energies of those flares follow a flare frequency distribution (FFD) where the most energetic flares occur far more rarely than lower energy flares \citep{Lacy1976}. FFDs have been found to depend on stellar spectral type (a proxy for mass and surface temperature) as well as other indicators of stellar activity due to a dependence on the underlying magnetic field strength \citep{Hilton2011phd,Hawley2014}. The dynamo that generates stellar magnetic fields (and thus flares) depends strongly on the rotation rate of the stars, both above and below the fully convective boundary \citep{Browning2006,Browning2008}. As stars age, their rotational energy is dissipated through a combination of a steady stellar winds and transient magnetic emission events, leading to weaker magnetic fields and net lower magnetic activity. A well-measured flare frequency distribution across a range of masses, ages, and metallicities would be an important step in understanding the magnetic evolution of M dwarfs and other main sequence stars. 

Powerful flares can also present a challenge to the surface habitability of exoplanets that orbit M dwarf hosts. The habitable zones (where liquid water can persist on the surface of a rocky planet) of M dwarfs are 10--100 times closer to their host star than that of the Sun \citep{Luger2015}, so the UV emission observed from most flares will have a stronger effect on planets around M dwarfs. UV emission, if too intense, can first deplete the outer ozone layer and then pass through an exoplanet atmosphere to destroy DNA and other essential molecules. On early- to mid-M dwarfs, UV emission alone is not strong enough to hinder life dramatically, either during quiescence \citep{France2016} or during a single flare \citet{Segura2010}. On later M dwarfs \citep{OMalley-James2017} or M dwarfs with a high rate of large flares \citep{Tilley2017}, the UV emission from those flares may play a role in depleting the upper ozone of an oxygen atmosphere but is unlikely to destroy it completely. 

The observed UV emission from flares may not prevent growth of life around M dwarfs, but powerful flares are often accompanied by energetic particles, either emitted as a particle beam or released as part of a coronal mass ejection (CME); these CMEs could have dramatic effects on the atmospheres of otherwise habitable planets \citep{Khodachenko2007,Lammer2007}. Outside our Solar system, there are no observations of particle beams and only a handful of observations that could be consistent with CMEs \citep[e.g.,][]{Vida2016}, but scaling from correlations between Solar flares and these energetic ejections indicates that M dwarf planets would be hit by strong particle beams multiple times per day \citep{Kay2016,Youngblood2017} and that those events could work together to deplete the ozone \citep{Tilley2017}. Because these results rely on extrapolating known flare rates to estimate particle events, it is essential to determine flare rates as a function of stellar mass and age to calculate the effect on a wider range of planets orbiting M dwarfs over a large range of spectral types and ages. 

The \textit{Kepler} mission \citep{Borucki2010} and its extension into K2 \citep{Howell2014} have proven to be an essential resource for flare observations at optical wavelengths. While the relatively red pass-band of the \textit{Kepler} space craft is not ideal for observing the blue/UV wavelengths where the flare and photosphere contrast is highest, the precision and cadence ($\sim$1 minute) of the photometry has enabled the detection of large numbers of stellar flares. Results from \textit{Kepler} include catalogs of flares across the main sequence \citep{Davenport2016,Walkowicz2011}, superflares ($E>10^{33}$~ergs) on Solar-type stars \citep{Maehara2012}, detailed analyses of the many M dwarf flares \citep{Hawley2014,Ramsay2015,Silverberg2016}, and the first observations of white-light flares on an L dwarf \citep{Gizis2013}. While data from \textit{Kepler} are ideal for detailed time-series analysis of flares, the mission is limited due to the need to pre-select a limited number of targets in each quarter and field. Most stars are observed because they are likely to be particularly good targets for exoplanet searches, and the Kepler/K2 stars are not representative of the underlying population of M dwarfs. To assess the complete flare rate across an unbiased sample of stars, a different approach is needed. 

The All-Sky Automated Survey of Supernovae (ASAS-SN; \citealp{Shappee2014}) is a survey with the main objective of detecting the nearest, brightest supernovae in the Universe \citep{Shappee2016a,Holoien2017a,Holoien2017b,Holoien2017c}, that has proven to be a valuable resource for a variety of transient events both inside and outside our Galaxy, including the three brightest optical tidal disruption events \citep{Holoien2014,Holoien2016,Holoien2016b}, the most luminous SN \citep{Dong2016}, AGN flares \citep{Shappee2014}, dramatic variability on young stars \citep{Holoien2014,Herczeg2016}, novae \citep{li2017}, and cataclysmic variables \citep{Davis2015}. ASAS-SN has already found very strong flares on M dwarfs \citep[e.g.,][]{Stanek2013,Schmidt2014,Simonian2016,Rodriguez2018} and the strongest L dwarf flare \citep{Schmidt2016a}. Because of its relatively bright magnitude range, flaring M dwarfs detected with ASAS-SN are primarily bright, nearby M dwarfs that can be examined in detail more easily than those in deeper surveys like LSST or Pan-STARRS.  These flare detections are thus useful both for understanding nearby M dwarfs and for classifying flares in future time-domain surveys.  

We describe the selection of flare candidates in ASAS-SN (Section~\ref{sec:asassn}) and our follow-up observations (Section~\ref{sec:obs}). We then characterize the flares in Section~\ref{sec:flareprop} and their stars in Section~\ref{sec:stellarprop}. In Section~\ref{sec:indiv}, we discuss individual objects of interest and we summarize our results in Section~\ref{sec:summary}. 

\section{ASAS-SN flare detections and sample selection}
\label{sec:asassn}
The ASAS-SN survey is a long-term project to monitor the entire sky for bright transient objects with as fast a cadence as possible.  ASAS-SN began taking data in November 2011 with two cameras in Hawaii and in April 2013 began searching for transients in real time. Since 2013, ASAS-SN has expanded multiple times, as detailed by \citet{Holoien2017b} and \citet{Kochanek2017}. The data for the M dwarfs examined here were taken while ASAS-SN was taking data only at the original two sites at Hawaii and Chile.  Each ASAS-SN unit now consists of four 14-cm aperture Nikon telephoto lenses on a common mount allowing the two ASAS-SN units used in this study to cover $\sim$16,000~deg$^2$ every clear night to $V \approx 16.5-17.3$~mag, depending on lunation. This allowed us to scan the visible (Sun constrained) sky every two nights.  ASAS-SN currently has five robotic units distributed adding units to sites in Chile, Texas, and South Africa and we will add a sixth unit in China by the end of 2018.  We currently observe the entire visible sky every 20 hours to $g \approx 18.5$~mag.

The telescopes are scheduled automatically and the data are reduced in real-time.  Our transient detection pipeline typically identifies new transient sources less than an hour after the data are taken.  Candidates are then scanned by team members in North America, South America, Asia and Europe to determine if the sources are likely real, and to provide a rough classification based on public survey photometry, literature measurements where available, and the previous ASAS-SN light curve at that location. Transient events that are determined to be real are immediately made public on the ASAS-SN transients page\footnote{\url{http://www.astronomy.ohio-state.edu/asassn/transients.html}}. SNe and CV outbursts are the most common detections, including 500+ and 1250+ candidates as of the end of 2016. 

When we first retrieved our target list on 2016 April 6, the ASAS-SN transient list included 1518 total transients, 48 of which were flagged as possible M dwarf flares. Of those, five were published in Astronomer's Telegrams \citep{Stanek2013,Stanek2013a,Holoien2014a,Shappee2016,Simonian2016}, and two had been examined in detail \citep{Schmidt2014,Schmidt2016a}. To perform a check on the completeness of the  identification of ASAS-SN transient events as flares on M dwarfs, we queried the Sloan Digital Sky Survey \citep[SDSS][]{Aihara2011,York2000}, Two Micron All Sky Survey \citep[2MASS][]{Skrutskie2006}, and Wide-field Infrared Survey Explorer \citep[WISE][]{Wright2010} photometric databases. We color-selected objects at the positions of ASAS-SN transients that were likely to be M dwarfs based on their $i-z$ colors (when SDSS photometry was available) or their 2MASS and WISE colors (when SDSS photometry was not available). SDSS and 2MASS photometry were subject to the color and quality cuts described in \citet{Schmidt2015}, and $rizK_S$ magnitudes are included in Table~\ref{tab:prop}. We recovered 25 of the 48 candidates and identified an additional 11 candidates.  The 23 transient events originally identified as flares on M dwarfs but not recovered in our search were either missing some data in the ASAS-SN transients page or were flagged for faintness or saturation in one of the surveys. The additional 11 candidates recovered were very similar to M dwarfs, but were classified by ASAS-SN as CV outbursts and/or had red $J-W2$ colors consistent with young accreting stars. We retained the brightest two of those objects (ASASSN-15ep and ASASSN-15af) as a check on the initial classifications. 

\renewcommand{\arraystretch}{0.9}
\begin{deluxetable*}{llllllllll} \tablewidth{0pt} \tabletypesize{\scriptsize} 
\tablecaption{Properties of ASAS-SN Flare Dwarfs \label{tab:prop} }
\tablehead{ \colhead{ASAS-SN}  &  \colhead{Spectral} & \colhead{$H\alpha$}  & \colhead{} & \colhead{} & \colhead{} & \colhead{} & \colhead{} & \colhead{} & \colhead{}  \\ 
\colhead{ID}  &  \colhead{Type} & \colhead{EW (\AA)}  & \colhead{$V$} & \colhead{$R$} & \colhead{$I$} & \colhead{$r$} & \colhead{$i$} & \colhead{$z$} & \colhead{$K_S$}  } 
\startdata 
13ba & M6 & $8.4\pm0.1$ & $17.42\pm0.01$ & $15.91\pm0.02$ & $14.21\pm0.01$ & $16.76\pm0.02$ & \nodata & $13.99\pm0.02$ & $11.45\pm0.02$   \\
13be & M4 & $3.2\pm1.2$ & $18.40\pm0.01$ & $17.16\pm0.01$ & $15.58\pm0.04$ & $17.83\pm0.01$ & $16.31\pm0.01$ & $15.48\pm0.03$ & $13.31\pm0.04$   \\
13bf & M2 & $2.7\pm0.1$ & $17.22\pm0.01$ & $16.19\pm0.01$ & $15.12\pm0.01$ & $16.60\pm0.01$ & $15.70\pm0.03$ & $15.21\pm0.03$ & $13.10\pm0.04$   \\
13bg & M6 & $3.6\pm0.2$ & $17.13\pm0.01$ & $15.53\pm0.01$ & $13.91\pm0.01$ & $16.51\pm0.01$ & $14.61\pm0.02$ & $13.54\pm0.02$ & $10.95\pm0.02$   \\
13bh & M5 & $5.8\pm0.8$ & $16.71\pm0.05$ & $15.53\pm0.02$ & $13.88\pm0.03$ & \nodata & \nodata & \nodata & $11.70\pm0.02$   \\
13bi & M4 & $1.3\pm0.1$ & $18.08\pm0.01$ & $16.67\pm0.01$ & $15.11\pm0.01$ & $17.40\pm0.01$ & $15.87\pm0.01$ & $15.02\pm0.01$ & $12.69\pm0.03$   \\
13bk & M6 & $29.1\pm2.1$ & $18.68\pm0.01$ & $17.36\pm0.02$ & $15.65\pm0.01$ & $18.06\pm0.02$ & $16.43\pm0.02$ & $15.54\pm0.02$ & $13.17\pm0.04$   \\
13bl & M3 & $5.6\pm0.1$ & $14.83\pm0.01$ & $13.63\pm0.01$ & $12.18\pm0.01$ & $14.15\pm0.01$ & \nodata & $12.14\pm0.02$ & $9.98\pm0.02$   \\
13bn & M4 & $5.0\pm0.3$ & $17.16\pm0.02$ & $15.88\pm0.01$ & $14.39\pm0.02$ & $16.48\pm0.01$ & $15.06\pm0.01$ & $14.28\pm0.01$ & $11.94\pm0.02$   \\
13bt & M7 & $4.5\pm0.0$ & $19.14\pm0.06$ & $17.55\pm0.06$ & $15.67\pm0.01$ & $18.68\pm0.02$ & $16.55\pm0.01$ & $15.42\pm0.02$ & $12.78\pm0.03$   \\
13cb & M8 & $26.8\pm0.5$ & $21.83\pm0.40$ & \nodata & \nodata & $21.23\pm0.05$ & $18.65\pm0.02$ & $17.08\pm0.02$ & $13.91\pm0.05$   \\
13cm & M4 & $6.1\pm2.3$ & $16.21\pm0.03$ & \nodata & $13.23\pm0.04$ & $15.41\pm0.02$ & \nodata & $13.07\pm0.02$ & $10.74\pm0.02$   \\
13cr & M4 & $7.0\pm1.5$ & $14.82\pm0.02$ & $13.63\pm0.01$ & $12.20\pm0.01$ & \nodata & \nodata & \nodata & $9.69\pm0.02$   \\
13de & M4 & $7.4\pm1.0$ & $18.14\pm0.02$ & \nodata & $15.18\pm0.03$ & $17.57\pm0.02$ & $16.03\pm0.02$ & $15.20\pm0.02$ & $12.84\pm0.03$   \\
13di & M5 & $5.1\pm0.1$ & $19.97\pm0.06$ & $18.78\pm0.05$ & $17.05\pm0.06$ & \nodata & \nodata & \nodata & $14.80\pm0.11$   \\
13dj & M2 & $<$0.75 & $23.09\pm0.04$ & $21.80\pm0.02$ & $20.49\pm0.02$ & $22.59\pm0.18$ & $21.22\pm0.08$ & $20.20\pm0.14$ & \nodata   \\
13dk & M5 & $7.8\pm3.9$ & $19.91\pm0.02$ & $18.37\pm0.02$ & $16.74\pm0.02$ & $19.17\pm0.02$ & $17.53\pm0.02$ & $16.63\pm0.02$ & $14.33\pm0.05$   \\
14bj & M4 & $2.6\pm0.6$ & $19.73\pm0.01$ & $18.32\pm0.01$ & $16.71\pm0.01$ & $19.13\pm0.01$ & $17.52\pm0.02$ & $16.63\pm0.02$ & $14.34\pm0.07$   \\
14bm & M5 & $6.5\pm0.8$ & $20.47\pm0.01$ & $18.95\pm0.01$ & $17.28\pm0.01$ & $19.75\pm0.02$ & $18.07\pm0.02$ & $17.15\pm0.02$ & $14.62\pm0.10$   \\
14bn & M4 & $4.5\pm1.2$ & $17.40\pm0.04$ & \nodata & $14.23\pm0.01$ & \nodata & \nodata & \nodata & $12.16\pm0.02$   \\
14cx & M6 & $8.2\pm0.8$ & $19.62\pm0.02$ & $17.96\pm0.02$ & $16.24\pm0.03$ & \nodata & \nodata & \nodata & $13.50\pm0.03$   \\
14dv & M6 & $10.3\pm0.7$ & $20.72\pm0.01$ & $18.70\pm0.08$ & $17.15\pm0.09$ & \nodata & \nodata & \nodata & $14.60\pm0.08$   \\
14ea & M5 & $9.5\pm0.1$ & $15.78\pm0.01$ & $14.38\pm0.01$ & $13.01\pm0.01$ & $15.22\pm0.01$ & \nodata & $12.61\pm0.02$ & $10.20\pm0.02$   \\
14fi & M5 & $5.6\pm1.9$ & $17.21\pm0.03$ & \nodata & $14.11\pm0.04$ & \nodata & \nodata & \nodata & $11.82\pm0.02$   \\
14gj & M5 & $8.5\pm0.6$ & $20.63\pm0.05$ & $18.95\pm0.03$ & $17.05\pm0.04$ & \nodata & \nodata & \nodata & $14.58\pm0.08$   \\
14gn & M5 & $6.7\pm1.3$ & $21.38\pm0.04$ & $19.68\pm0.02$ & $17.98\pm0.02$ & $20.57\pm0.04$ & $18.76\pm0.02$ & $17.78\pm0.02$ & $15.03\pm0.13$   \\
14hc & K5 & $<$0.75 & $18.99\pm0.02$ & $17.59\pm0.02$ & $15.80\pm0.03$ & $18.30\pm0.01$ & $16.59\pm0.01$ & $15.65\pm0.01$ & $13.26\pm0.03$   \\
14hz & M6 & $6.9\pm1.0$ & $20.16\pm0.02$ & $18.66\pm0.08$ & $17.16\pm0.12$ & \nodata & \nodata & \nodata & $14.56\pm0.08$   \\
14ji & M7 & $10.1\pm5.0$ & $22.42\pm0.05$ & $20.35\pm0.02$ & $18.41\pm0.01$ & $21.54\pm0.07$ & $19.30\pm0.02$ & $17.99\pm0.03$ & $15.05\pm0.14$   \\
14jw & M7 & $6.4\pm0.2$ & $18.96\pm0.06$ & $17.34\pm0.04$ & $15.46\pm0.04$ & \nodata & \nodata & \nodata & $12.36\pm0.03$   \\
14jy & M4 & $4.8\pm0.3$ & $15.05\pm0.01$ & \nodata & \nodata & \nodata & \nodata & \nodata & $10.12\pm0.02$   \\
14ke & M8 & $16.4\pm7.3$ & $21.03\pm0.04$ & $19.38\pm0.07$ & $17.60\pm0.03$ & \nodata & \nodata & \nodata & $14.33\pm0.08$   \\
14lc & M9 & $20.0\pm0.3$ & $24.20\pm0.10$ & $21.44\pm0.03$ & $19.29\pm0.02$ & \nodata & \nodata & \nodata & $15.17\pm0.12$   \\
14mz & M6 & $5.0\pm0.0$ & $19.38\pm0.02$ & $17.68\pm0.02$ & $15.82\pm0.03$ & $18.73\pm0.02$ & $16.73\pm0.02$ & $15.64\pm0.02$ & $13.07\pm0.03$   \\
15ep & M5 & $6.8\pm2.8$ & $19.66\pm0.03$ & \nodata & $16.40\pm0.05$ & \nodata & \nodata & \nodata & $13.88\pm0.06$   \\
15kl & M3 & $2.7\pm0.4$ & $17.21\pm0.04$ & $15.92\pm0.06$ & $14.29\pm0.07$ & \nodata & \nodata & \nodata & $11.73\pm0.02$   \\
15ll & M4 & $<$0.75 & $22.60\pm0.03$ & $21.34\pm0.01$ & $19.91\pm0.01$ & $22.12\pm0.09$ & $20.75\pm0.04$ & $19.82\pm0.08$ & \nodata   \\
15oy & M8 & $9.2\pm0.3$ & $18.63\pm0.01$ & $16.54\pm0.01$ & $14.55\pm0.01$ & $23.73\pm1.23$ & $24.36\pm61.59$ & $21.84\pm3.38$ & $11.26\pm0.02$   \\
15tv & M5 & $6.6\pm0.0$ & $19.66\pm0.01$ & $18.24\pm0.02$ & $16.44\pm0.01$ & $18.96\pm0.02$ & $17.21\pm0.02$ & $16.27\pm0.02$ & $13.83\pm0.04$   \\
16ae & L0 & $26.4\pm0.2$ & $23.05\pm0.43$ & \nodata & $19.43\pm0.03$ & \nodata & $20.75\pm0.05$ & $18.87\pm0.04$ & $15.47\pm0.21$   \\
16cx & M9 & $6.3\pm0.2$ & $22.67\pm0.42$ & \nodata & \nodata & $22.07\pm0.12$ & $19.30\pm0.02$ & $17.67\pm0.02$ & $14.68\pm0.09$   \\
16di & M7 & $6.6\pm2.9$ & $20.36\pm0.01$ & $18.54\pm0.01$ & $16.72\pm0.01$ & $19.68\pm0.02$ & $17.59\pm0.01$ & $16.35\pm0.02$ & $13.48\pm0.05$   \\
16dj & M4 & $6.1\pm0.2$ & $16.03\pm0.01$ & $14.76\pm0.01$ & $13.16\pm0.01$ & $15.30\pm0.02$ & $13.85\pm0.02$ & $13.01\pm0.02$ & $10.81\pm0.02$   \\
16dr & M5 & $6.1\pm1.4$ & $19.60\pm0.02$ & $17.84\pm0.03$ & $15.91\pm0.04$ & $18.96\pm0.02$ & $16.84\pm0.01$ & $15.70\pm0.01$ & $13.03\pm0.03$   \\
16du & M6 & $5.6\pm0.1$ & $21.80\pm0.07$ & $20.37\pm0.02$ & $18.37\pm0.01$ & $21.34\pm0.06$ & $19.16\pm0.02$ & $18.08\pm0.03$ & $15.42\pm0.16$   \\
16gt & M6 & $4.9\pm0.4$ & $17.09\pm0.02$ & \nodata & $13.56\pm0.05$ & \nodata & \nodata & \nodata & $11.12\pm0.02$   \\
16hq & M8 & $9.3\pm1.7$ & $21.55\pm0.02$ & $19.27\pm0.01$ & $17.13\pm0.01$ & \nodata & \nodata & \nodata & $13.64\pm0.04$   \\
16kq & M4 & $4.5\pm0.9$ & $18.17\pm0.01$ & \nodata & $15.10\pm0.01$ & $17.44\pm0.01$ & $15.92\pm0.01$ & $15.08\pm0.01$ & $12.81\pm0.03$   \\
GJ3039 & M4 & $5.0\pm0.1$ & $12.66\pm0.01$ & \nodata & \nodata & \nodata & \nodata & \nodata & $11.82\pm0.02$   \\
\enddata
\end{deluxetable*}
\renewcommand{\arraystretch}{1.0}

We also added six additional candidate flaring M dwarfs that were observed between 2016 April and 2016 October, for a total list of 55 candidate flaring M dwarfs. Of those, two (ASASSN-13cn and ASASSN-14bk) were not observed because they fell outside color cuts designed to limit the sample to M dwarfs. Our final target list of 53 objects is given in Table~\ref{tab:obs}. 


\renewcommand{\arraystretch}{0.9}
\begin{deluxetable*}{llllll} \tablewidth{0pt} \tabletypesize{\scriptsize}
\tablecaption{Observation Log \label{tab:obs} }
\tablehead{ \colhead{ASAS-SN}  & \colhead{} & \colhead{} & \multicolumn{2}{c}{Spectroscopy} & \colhead{Photometric} \\
\colhead{Number} & \colhead{RA} & \colhead{dec} & \colhead{Telescope/Instr} & \colhead{ET (s)} & \colhead{Bands} } 
\startdata
13ba & 01 17 00.6 & -05 06 04.6 & Magellan/Baade & 2 x 300      & $VRI$ \\
13be & 18 04 08.7 & +46 56 04.0 & MDM            & 4 x 1800     & $VRI$ \\
13bf & 23 06 12.9 & -09 42 33.2 & Magellan/Baade & 2 x 300      & $VRI$ \\
13bg & 23 54 22.4 & +07 02 48.7 & Magellan/Baade & 1 x 450      & $VRI$ \\
13bh & 01 35 13.7 & +18 23 13.9 & MDM            & 3 x 1200     & $VRI$ \\
13bi & 14 58 05.5 & +02 44 33.6 & Magellan/Baade & 2 x 600      & $VRI$ \\
13bk & 17 08 56.9 & +71 51 48.1 & MDM            & 4 x 1800     & $VRI$ \\
13bl & 17 28 39.7 & +28 25 35.4 & MDM            & 2 x 900      & $VRI$ \\
13bn & 23 54 03.3 & +37 56 52.6 & MDM            & 3 x 1200     & $VRI$ \\
13bt & 15 53 26.1 & +48 08 59.4 & LBT            & 2 x 6 x 800  & $VRI$ \\
13cm & 01 46 51.4 & -16 52 19.7 & du Pont        & 1 x 1500     & $VI$  \\
13cr & 04 36 49.2 & -02 49 29.7 & MDM            & 2 x 1200     & $VRI$ \\
13de & 04 16 23.4 & -05 17 27.9 & du Pont        & 4 x 1800     & $VI$ \\ 
     &            &             & MDM            & 1 x 1800     & \\
13di & 07 14 51.8 & +64 27 02.2 & LBT            & 2 x 4 x 750  & $VRI$ \\
13dj & 01 38 15.5 & +16 39 43.1 & Magellan/Baade & 7 x 1800     & $VRI$ \\
13dk & 09 46 23.8 & +19 55 10.6 & Magellan/Baade & 1 x 1500 + 1 x 1800  & $VRI$ \\
14aa & 09 18 51.4 & +19 27 33.6 & Magellan/Baade & 3 x 2500     &  $VRI$ \\
14be & 11 31 24.9 & +04 41 16.5 & Magellan/Baade & 5 x 2500     & $VRI$ \\
14bj & 14 49 37.7 & +16 56 56.9 & Magellan/Baade & 2 x 1200     & $VRI$ \\
14bm & 14 22 13.3 & +03 03 04.2 & Magellan/Baade & 2 x 1200     & $VRI$ \\
14bn & 01 22 42.7 & -29 11 18.4 & du Pont        & 5 x 1800     & $VI$ \\
14cx & 18 04 00.0 & +20 56 46.2 & Magellan/Baade & 2 x 750      & $VRI$ \\
14dv & 01 50 38.4 & -24 23 56.4 & Magellan/Baade & 2 x 1200     & $VRI$ \\
14ea & 21 55 17.5 & -00 45 49.3 & Magellan/Baade & 2 x 300      & $VRI$ \\
14fi & 02 03 32.9 & -67 31 26.8 & du Pont        & 3 x 1500     & $VI$ \\
14gj & 03 12 15.7 & +26 04 40.6 & Magellan/Baade & 3 x 800      & $VRI$ \\
14gn & 23 10 11.1 & +24 02 17.1 & Magellan/Baade & 3 x 1200     & $VRI$ \\
14hc & 23 55 51.1 & +24 48 51.9 & MDM            & 6 x 1800     & $VRI$ \\
14hz & 20 45 26.6 & -34 03 39.9 & Magellan/Baade & 2 x 1200     & $VRI$ \\
14ji & 04 05 51.4 & -11 19 17.0 & Magellan/Baade & 4 x 1500     & $VRI$ \\
14jw & 21 47 18.3 & -41 16 20.8 & Magellan/Baade & 2 x 600      & $VRI$ \\
14jy & 07 06 58.9 & -62 21 10.9 & du Pont        & 2 x 750      & $VI$ \\
14ke & 02 19 08.2 & +11 07 45.9 & Magellan/Baade & 3 x 1200     & $VRI$ \\
14lc & 12 02 29.6 & +24 12 12.4 & LBT            & 2 x 4 x 1500 & $VRI$ \\
14mz & 08 51 13.9 & +19 12 21.5 & LBT            & 2 x 3 x 800  & $VRI$ \\
15af & 01 57 54.9 & -54 30 37.9 & Magellan/Baade & 1 x 600      & $VRI$ \\
     &            &             & du Pont        & 1 x 1800     & \\
     &            &             & du Pont        & 1 x 1800     & \\
15ep & 08 21 06.2 & -72 20 11.2 & du Pont        & 6 x 1800     & $VI$ \\
15kl & 16 06 14.8 & -04 35 49.7 & Magellan/Baade & 2 x 300      & $VRI$ \\
15ll & 22 40 02.7 & +26 30 45.1 & Magellan/Baade & 4 x 1800     & $VRI$ \\
     &            &             & LBT            & 2 x 4 x 1200 & $VRI$ \\
15oy & 02 48 35.7 & +19 16 26.6 & Magellan/Baade & 2 x 1000     & $VRI$ \\
15tv & 01 35 11.7 & +26 25 38.8 & LBT            & 2 x 3 x 800  & $VRI$ \\
16cx & 16 11 58.5 & +54 56 42.9 & LBT            & 2 x 2 x 1200 & \nodata \\
16di & 14 16 36.4 & +01 52 05.7 & Magellan/Baade & 2 x 1200     & $VRI$ \\
16dj & 10 07 17.7 & +69 20 46.2 & MDM            & 4 x 1200     & $VRI$ \\
16dr & 12 27 04.2 & +25 41 01.6 & Magellan/Baade & 3 x 1800     & $VRI$ \\
16du & 07 53 11.6 & +28 16 42.4 & LBT            & 2 x 6 x 1200 & $VRI$ \\
16gt & 10 37 02.4 & -18 27 44.8 & du Pont        & 2 x 1500     & $BVI$ \\
16hl & 16 04 55.2 & -72 23 18.3 & du Pont        & 3 x 1500     & $VI$ \\
16hq & 16 05 16.5 & +00 07 05.5 & Magellan/Baade & 3 x 1200     & $VRI$ \\
16kq & 08 09 33.9 & +02 15 39.7 & du Pont        & 4 x 1800 + 1 x 1500 & $BVI$ \\
GJ3039 & 00 32 34.9 & +07 29 26.5 & du Pont      & 2 x 300      & $V$ \\
\enddata
\end{deluxetable*}
\renewcommand{\arraystretch}{1.0}

\section{Observations}
\label{sec:obs}
We obtained follow-up optical photometry and spectroscopy to characterize the host stars of the 53 candidate M dwarfs and flare events detected by ASAS-SN. 

\subsection{Photometry}
We obtained follow-up $VRI$ photometry for the majority of our sources, both to aid in source classification (combined with survey photometry) and to provide quiescent $V$-band measurements to compare to the $V$-band flare detections. Typically, the photometry was obtained when acquiring sources for spectroscopic observations. We obtained optical imaging using the Inamori-Magellan Areal Camera \& Spectrograph (IMACS; \citealp{Dressler2011}) on the Baade-Magellan~6.5-m telescope, the Ohio State Multi-Object Spectrograph (OSMOS; \citealp{Martini2011}) on the MDM Observatory Hiltner 2.4-m telescope, and the Wide Field Reimaging CCD Camera (WFCCD) on the Ir\'{e}n\'{e}e du Pont 100-inch Telescope.  All data were reduced with standard routines in the IRAF {\tt ccdred} package.  We performed aperture photometry on the reduced images using the IRAF {\tt apphot} package.  The data were calibrated using Sloan Digital Sky Survey (SDSS; \citealt{York2000}) Data Release 7 \citep{Abazajian2009} data when available and using the AAVSO Photometric All-Sky Survey \citep[APASS;][]{Henden2014} when SDSS was unavailable. When using SDSS, the Sloan filters were transformed onto the Johnson-Cousins magnitude system using transformations presented by Robert Lupton\footnote{\url{http://www.sdss.org/dr5/algorithms/sdssUBVRITransform.html}}.  The resulting photometry is presented in Table~\ref{tab:prop}. 

We initially identified flares based on quick photometry from ASAS-SN images. To examine the lightcurves in more detail, we performed aperture photometry at the location of each flare as described in \citet{Kochanek2017}.  Briefly, the photometry is done using the IRAF apphot package and calibrated to APASS \citep{Henden2014}. However, for ASASSN-13be there is a nearby blended star and so we performed aperture photometry on subtracted images, calibrated to APASS, and then added back in the quiescent flux as measured from our higher resolution OSMOS images. We also examined the ASAS-SN light curve for each source to look for additional flares and to characterize the flares detected by ASAS-SN (see Section \ref{sec:flareprop}). In addition to the original flares that triggered an ASAS-SN classification, we detected six additional significant ($\Delta V < -1.0$) flares on five stars, which we include in our analysis. The photometry of each flare along with 3-sigma upper limits for images within $\sim$24 hours of each flare event are presented in Table~\ref{tab:flarephot}. 

\renewcommand{\arraystretch}{0.9}
\begin{deluxetable}{llll} \tablewidth{0pt} \tabletypesize{\scriptsize} 
\tablecaption{ASAS-SN Photometry of Flares \label{tab:flarephot} }
\tablehead{ \colhead{ASAS-SN}  &  \colhead{UT} & \multicolumn{2}{C}{$V$}  \\ 
\colhead{number}  &  \colhead{date} & \colhead{3$\sigma$ limit}  & \colhead{detection}  } 
\startdata 
13ba & 2013-06-28.5679994 & 16.71 & \nodata \\
            & 2013-06-28.5693679 & 16.54 & \nodata \\
            & 2013-07-02.5574708 & 16.22 & 12.04$\pm$0.01 \\
            & 2013-07-02.5588631 & 16.39 & 12.40$\pm$0.01 \\
            & 2013-07-05.5604776 & 17.53 & \nodata \\
            & 2013-07-05.5618807 & 17.50 & \nodata \\
13be & 2013-07-05.4018670 & 18.04 & \nodata \\
            & 2013-07-05.4032605 & 18.22 & 18.19$\pm$0.35 \\
            & 2013-07-06.4012965 & 17.98 & 16.02$\pm$0.06 \\
            & 2013-07-06.4026820 & 18.08 & 16.25$\pm$0.07 \\
            & 2013-07-08.3802245 & 17.87 & \nodata \\
\enddata
\tablecomments{Example table provided for guidance; full table available online.}
\end{deluxetable}
\renewcommand{\arraystretch}{1.0}

\subsection{Spectroscopy}
To confirm the stellar nature of the M dwarf candidates and to characterize them in terms of spectral type and quiescent chromospheric activity (through the H$\alpha$ emission line), we obtained low-resolution (R$\sim$2000--4000) optical (6000--8000\AA) spectra from a variety of telescopes. The instruments and telescopes used for each target are listed in Table~\ref{tab:obs}. We adopted the spectra for ASASSN-13cb and ASASSN-16ae described by \citet{Schmidt2014} and \citet{Schmidt2016a}, respectively. 

We obtained optical spectroscopy for bright southern sources on the Ir\'{e}n\'{e}e du Pont telescope at the Las Campanas Observatory (LCO). We used the Wide Field CCD in long slit grism mode yielding a wavelength range $370-910$ nm, and resolution 8\AA. We obtained optical spectroscopy for faint southern sources using the Inamori-Magellan Areal Camera \& Spectrograph (IMACS; \citealp{Dressler2011}) on the Baade-Magellan~6.5-m telescope.  We observed candidates with the F/2 camera, 0\farcs7-0\farcs9 slits, 300 lines per mm grism at a blaze angle of 17.5 degrees yielding a wavelength range $390-800$ nm and a dispersion of $\sim 1.3$ \AA{} pixel$^{-1}$. Optical spectroscopy for bright northern sources was obtained using the the Ohio State Multi-Object Spectrograph (OSMOS; \citealp{Martini2011}) on the MDM Observatory Hiltner 2.4-m telescope.  We observed candidates with 0\farcs9-1\farcs2 slits and VPH grism yielding a wavelength range $390-680$ nm and R$\sim1600$. 

We reduced the spectra from Magellan, du Pont, and MDM using standard IRAF routines, supplemented by cosmic ray removal using the LA cosmic detection algorithm \citep{van-Dokkum2001}. Wavelength calibration for Magellan and DuPont used arc lamps taken at the position of the observations and resulted in solutions good to 0.1--0.3 pixels. For MDM, arc lamps were typically taken once per night, resulting in wavelength solutions good to 0.5 pixels with 1--4~pixel offsets that we corrected using the H$\alpha$ emission line. No radial velocity standards were taken with any telescope, so measuring velocities from the spectra was not attempted. For all three telescopes, we performed flux calibration using observations of a single flux standard per night.

We obtained spectra for the faintest northern sources with the Multi-Object Double Spectrograph 1 \citep[MODS1;][]{Pogge2010} mounted on the dual 8.4-m Large Binocular Telescope (LBT) on Mt. Graham. The MODS data were reduced with a combination of the  \textsc{modsccdred}\footnote{\url{http://www.astronomy.ohio-state.edu/MODS/ Software/modsCCDRed/}} \textsc{python} package and the \textsc{modsidl}  pipeline\footnote{\url{http://www.astronomy.ohio-state.edu/MODS/Software/modsIDL/}}.  The reductions included bias subtraction, flat-fielding, 1-D spectral extraction, wavelength calibration using an arc lamp, and flux calibration using a spectroscopic standard taken the same night.

We assigned optical spectral types to each star based on comparison to the \citet{Bochanski2007a} SDSS spectroscopic templates. Of the 53 stars, 47 were classified as M dwarfs with types M2--M9, as listed in Table~\ref{tab:prop}, along with one K5 dwarf (ASASSN-14he), one L0 dwarf \citep[ASASSN-16ae;][]{Schmidt2016a}. Spectral classifications for the stars are given in Table~\ref{tab:prop}, and photometry for the other objects is given in Table~\ref{tab:rej}. We include the L0 dwarf in our statistics and results because its chromospheric activity is not expected to be distinct from that of the 10 M7--M9 dwarfs included in the sample. We exclude the K5 dwarf because it is significantly more massive than the rest of the sample (types M2 and later).

In addition to these ASAS-SN flare stars, we also observed one previously identified TT Tauri star (see Section~\ref{sec:tt}) and one newly identified young star (see Section~\ref{sec:mys}). For one faint object, ASASSN-14aa, the spectrum of a $V=23.4$~mag source close to the ASAS-SN position is not stellar, and upon further examination the real source of the outburst is likely to be a $V=25.2$~mag point source nearby. Follow-up Magellan photometric observations of another faint source, ASASSN-14be, was unable to identify any source consistent with the ASAS-SN position.  We place 5-sigma limits of $V>25.6$, $R>24.9$, and $I>23.6$~mag at the position of ASASSN-14be, making its counterpart significantly too faint to obtain a spectrum of sufficient quality to provide a spectral type. Because the stellar photometry meets color selection criteria for M dwarfs, ASASSN-14aa and ASASSN-14be are likely to be $\Delta V <-10$~mag flares on distant M dwarfs. We do not include them in this analysis due to the lack of spectroscopic observations. 

\begin{deluxetable*}{lllllll} \tablewidth{0pt} \tabletypesize{\scriptsize}
\tablecaption{M dwarf Contaminants \label{tab:rej} }
\tablehead{ \colhead{ASAS-SN} & \colhead{other} & \multicolumn{5}{c}{Photometry}  \\
            \colhead{number}  & \colhead{designation}       & \colhead{$J$} & \colhead{$H$}           & \colhead{$K_S$}  & \colhead{$W1$} & \colhead{$W2$}} 
\startdata
ASASSN-15af & 2MASS J01575499-5430367 & 15.13$\pm$0.04 & 14.63$\pm$0.04 & 14.27$\pm$0.06 & 14.15$\pm$0.03 & 13.92$\pm$0.03\\ 
ASASSN-16hl & 2MASS J16045515-7223199 & 14.14$\pm$0.03 & 13.43$\pm$0.03 & 13.30$\pm$0.04 & 12.92$\pm$0.03 & 12.81$\pm$0.03\\ 
\enddata
\end{deluxetable*}

\section{Calculating Flare Energies}
\label{sec:flareprop}
The magnitudes for each observation of the 55 flares detected on 49 objects are given in Table~\ref{tab:flarephot}. Each ASAS-SN flare detection includes between one and six individual observations. Most of these observations span $\sim$2--8 minutes, with only three flares \citep[including ASASSN-13cb;][]{Schmidt2014} having additional ASAS-SN detections within less than a day. Typically, there are non-detections or detections at the quiescent (non-flaring) magnitude $\sim$24--48 hours before and after the event, but these data points do not provide useful constraints on flares, which typically last 1--6~hours \citep{Davenport2014}. To examine the data from each flare, we first convert both the quiescent and flaring $V$-band magnitudes to $V$-band fluxes, then subtract the quiescent flux from all flare data to obtain flare-only fluxes. 

Distance is an essential factor in converting flare fluxes to energies. We obtained distances (given in Table~\ref{tab:kine}) for 42 of the 49 stars via a cross-match to the \citet{Bailer-Jones2018} catalog of distances derived from Gaia DR2 parallaxes \citep{Gaia-Collaboration2018,Lindegren2018}. These Gaia distances have uncertainties of 1\% for the brighter ($V<19$) stars and closer to 7\% for the fainter stars. The remaining 7 stars include the nearby binary star GJ3039, the four faintest M dwarfs, and two stars that have poor astrometric solutions due to nearby background objects. We calculated distances for the remaining stars in our sample using available photometry. For four of the dwarfs, we combined our $V$-band photometry with 2MASS $K_S$-band to use the $V-K_S$ relation from \citet{Henry2004}. For the three dwarfs without reliable $K_S$ or Gaia matches, we used high quality SDSS photometry to calculate distances. For the faint early-M dwarfs ASASSN-13dj and ASASSN-15ll, we used the $r-z$ relation from \citet{Bochanski2011} to derive distances and for the L0 ASASSN-16ae, we used the $i-z$ relation from \citet[][in prep.]{Schmidt2018}. The photometric distances have a mean uncertainty of 18\%. Because we examined only the brightest flares, the ASAS-SN flare sample is biased towards nearby objects; the majority (43 of 49 objects) are located within 200~pc, with five additional objects between 200 and 400~pc, and only ASASSN-13dj and ASAS-SN15ll ($1090\pm220$~pc and $1000\pm190$~pc, respectively) found at larger distances.

\renewcommand{\arraystretch}{0.9}
\begin{deluxetable*}{l lll rr l} \tablewidth{0pt} \tabletypesize{\scriptsize}
\tablecaption{Kinematics \label{tab:kine} }
\tablehead{ \colhead{ASAS-SN} & \colhead{kinematics}  & \colhead{d [pc]} & \colhead{z [pc]} & \colhead{$\mu_{\alpha}\cos(\delta)$ } & \colhead{$\mu_{\delta}$} & \colhead{$V_{\rm tan}$} \\
            \colhead{number}  & \colhead{source}       & \colhead{}       & \colhead{}       & \colhead{[mas/yr]}          & \colhead{[mas/yr]}         & \colhead{km s$^{-1}$}    }
\startdata
13ba & Gaia & 113.4$\pm$2.1 & $-$89.4 & 154.8$\pm$0.3 & 67.7$\pm$0.1 & 91.4$\pm$1.7 \\ 
13be & Gaia & 127.1$\pm$1.3 & 73.1 & $-$131.7$\pm$0.1 & $-$142.1$\pm$0.2 & 117.5$\pm$1.2 \\ 
13bf & Gaia & 339.5$\pm$9.6 & $-$278.2 & $-$16.3$\pm$0.1 & $-$25.7$\pm$0.1 & 49.2$\pm$1.4 \\ 
13bg & Gaia & 52.9$\pm$0.4 & $-$27.3 & 23.2$\pm$0.2 & 97.6$\pm$0.1 & 25.3$\pm$0.2 \\ 
13bh & Gaia & 144.4$\pm$2.4 & $-$83.8 & $-$9.9$\pm$0.2 & 3.7$\pm$0.2 & 7.3$\pm$0.2 \\ 
13bi & Gaia & 106.3$\pm$0.9 & 97.7 & 8.0$\pm$0.1 & $-$21.7$\pm$0.1 & 11.7$\pm$0.1 \\ 
13bk & Gaia & 183.4$\pm$3.1 & 116.5 & $-$16.3$\pm$0.2 & 29.9$\pm$0.2 & 29.8$\pm$0.6 \\ 
13bl & Gaia & 43.9$\pm$0.1 & 36.6 & 11.5$\pm$0.1 & 51.5$\pm$0.1 & 11.1$\pm$0.0 \\ 
13bn & Gaia & 177.5$\pm$3.8 & $-$55.8 & $-$11.2$\pm$0.1 & 9.8$\pm$0.1 & 12.6$\pm$0.3 \\ 
13bt & Gaia & 171.7$\pm$18.8 & 144.7 & $-$45.6$\pm$1.1 & 44.9$\pm$1.4 & 52.4$\pm$5.9 \\ 
13cb & Gaia & 75.6$\pm$2.9 & $-$31.9 & 125.9$\pm$1.0 & $-$22.6$\pm$0.9 & 46.2$\pm$1.9 \\ 
13cm & Gaia & 59.1$\pm$0.2 & $-$41.7 & 54.0$\pm$0.1 & 16.6$\pm$0.1 & 15.9$\pm$0.1 \\ 
13cr & Gaia & 112.1$\pm$5.0 & $-$42.6 & 29.6$\pm$0.7 & $-$20.6$\pm$0.4 & 19.3$\pm$1.0 \\ 
13de & Gaia & 151.1$\pm$2.1 & $-$75.0 & 1.0$\pm$0.1 & $-$9.1$\pm$0.1 & 6.6$\pm$0.1 \\ 
13di & Gaia & 381.8$\pm$32.4 & 187.6 & $-$3.2$\pm$0.3 & $-$0.9$\pm$0.3 & 6.0$\pm$0.9 \\ 
13dj & r-z/2M-W & 1086.7$\pm$217.0 & $-$749.1 & \nodata & \nodata & \nodata \\ 
13dk & V-K/2M-W & 189.9$\pm$34.7 & 154.6 & 29.2$\pm$7.7 & $-$20.6$\pm$5.5 & 32.4$\pm$10.4 \\ 
14bj & Gaia & 185.3$\pm$6.7 & 176.7 & $-$8.3$\pm$0.4 & $-$44.0$\pm$0.3 & 39.6$\pm$1.5 \\ 
14bm & Gaia & 263.8$\pm$24.2 & 237.4 & $-$5.8$\pm$0.7 & $-$10.1$\pm$0.6 & 14.7$\pm$1.8 \\ 
14bn & Gaia & 83.0$\pm$0.6 & $-$67.3 & $-$76.9$\pm$0.2 & $-$6.6$\pm$0.1 & 30.5$\pm$0.2 \\ 
14cx & Gaia & 136.8$\pm$2.7 & 60.5 & 5.4$\pm$0.2 & $-$6.6$\pm$0.3 & 5.5$\pm$0.2 \\ 
14dv & Gaia & 176.6$\pm$11.7 & $-$156.6 & 54.1$\pm$0.6 & 20.5$\pm$0.4 & 48.8$\pm$3.3 \\ 
14ea & Gaia & 56.5$\pm$0.3 & $-$21.5 & 63.8$\pm$0.1 & $-$54.8$\pm$0.1 & 22.7$\pm$0.1 \\ 
14fi & Gaia & 123.5$\pm$0.8 & $-$77.0 & 16.2$\pm$0.1 & 23.7$\pm$0.1 & 16.9$\pm$0.1 \\ 
14gj & Gaia & 193.3$\pm$14.7 & $-$72.2 & $-$14.9$\pm$0.8 & $-$10.8$\pm$0.6 & 17.0$\pm$1.6 \\ 
14gn & Gaia & 199.8$\pm$17.1 & $-$94.7 & $-$34.7$\pm$0.8 & $-$24.6$\pm$0.6 & 40.5$\pm$3.6 \\ 
14hc & Gaia & 111.3$\pm$1.9 & $-$50.9 & $-$35.9$\pm$0.3 & $-$7.3$\pm$0.1 & 19.4$\pm$0.4 \\ 
14hz & Gaia & 186.3$\pm$11.0 & $-$98.4 & 18.4$\pm$0.5 & 5.4$\pm$0.4 & 17.0$\pm$1.2 \\ 
14ji & Gaia & 175.4$\pm$20.1 & $-$101.7 & $-$10.3$\pm$0.9 & $-$45.3$\pm$0.7 & 38.9$\pm$4.6 \\ 
14jw & Gaia & 47.9$\pm$0.4 & $-$21.7 & $-$198.4$\pm$0.3 & $-$126.3$\pm$0.4 & 53.8$\pm$0.5 \\ 
14jy & Gaia & 46.3$\pm$0.1 & $-$2.4 & $-$4.8$\pm$0.1 & $-$39.1$\pm$0.1 & 8.7$\pm$0.0 \\ 
14ke & Gaia & 142.6$\pm$24.7 & $-$87.8 & 60.7$\pm$1.1 & $-$13.3$\pm$0.9 & 42.3$\pm$7.4 \\ 
14lc & V-K/2M-W & 91.3$\pm$17.4 & 104.5 & 51.0$\pm$15.3 & 13.9$\pm$7.1 & 23.0$\pm$8.5 \\ 
14mz & V-K/2M-W & 74.4$\pm$13.5 & 57.5 & 68.9$\pm$3.7 & $-$9.9$\pm$7.8 & 24.7$\pm$5.4 \\ 
15ep & Gaia & 159.3$\pm$3.7 & $-$37.9 & $-$17.0$\pm$0.2 & 41.9$\pm$0.3 & 34.4$\pm$0.8 \\ 
15kl & Gaia & 65.8$\pm$0.5 & 51.0 & 74.5$\pm$0.2 & $-$44.1$\pm$0.1 & 27.2$\pm$0.2 \\ 
15ll & r-z/2M-W & 999.1$\pm$185.5 & $-$450.1 & \nodata & \nodata & \nodata \\ 
15oy & Gaia & 24.9$\pm$0.1 & 0.5 & 248.4$\pm$0.2 & $-$124.7$\pm$0.2 & 33.1$\pm$0.1 \\ 
15tv & Gaia & 148.3$\pm$4.2 & $-$70.8 & 36.2$\pm$0.4 & $-$14.1$\pm$0.3 & 27.5$\pm$0.8 \\ 
16ae & i-z/2M-W & 77.9$\pm$14.0 & $-$7.8 & 126.5$\pm$11.7 & 157.3$\pm$26.3 & 75.0$\pm$17.2 \\ 
16cx & Gaia & 79.7$\pm$2.8 & 70.9 & $-$20.4$\pm$1.2 & 75.3$\pm$0.8 & 29.6$\pm$1.2 \\ 
16di & Gaia & 137.2$\pm$10.6 & 130.7 & $-$10.9$\pm$1.0 & $-$19.2$\pm$0.7 & 14.4$\pm$1.4 \\ 
16dj & Gaia & 52.5$\pm$0.1 & 49.9 & $-$18.1$\pm$0.1 & 23.1$\pm$0.1 & 7.3$\pm$0.0 \\ 
16dr & Gaia & 100.0$\pm$4.0 & 114.5 & $-$13.6$\pm$0.4 & $-$5.9$\pm$1.1 & 7.1$\pm$0.6 \\ 
16du & Gaia & 237.6$\pm$48.5 & 116.3 & $-$2.1$\pm$1.2 & $-$24.9$\pm$0.6 & 28.3$\pm$6.0 \\ 
16gt & Gaia & 28.5$\pm$0.1 & 30.9 & $-$421.1$\pm$0.1 & $-$25.5$\pm$0.1 & 57.3$\pm$0.2 \\ 
16hq & Gaia & 108.7$\pm$5.4 & 79.1 & $-$51.9$\pm$0.8 & $-$14.6$\pm$0.6 & 28.0$\pm$1.5 \\ 
16kq & Gaia & 107.0$\pm$1.5 & 48.9 & $-$55.7$\pm$0.2 & 11.2$\pm$0.1 & 29.0$\pm$0.4 \\ 
GJ3039 & V-K/2M-W & 10.6$\pm$1.9 & 6.3 & $-$95.7$\pm$7.3 & 75.5$\pm$3.5 & 6.2$\pm$1.2 \\ 
\enddata
\end{deluxetable*}
\renewcommand{\arraystretch}{1.0}

\subsection{Lower Limits on $V$-band Flare Energies}
\label{sec:llenergy}
The survey strategy of ASAS-SN---acquiring two or three 90 second dithers on a field before moving on to another field---does not allow for a detailed analysis of flare lightcurves, but the data are sufficient to place accurate lower limits on the $V$-band flare energy. Given a single detection that indicates a flare-type brightening, the lowest energy flare that could correspond to that observation is a simple, classical flare with the peak of the flare occurring during that single observed point. We calculate the flare energy based on the fit of a simple, classical flare to the data and refer to it as a lower-limit because other light-curves that fit the points have a higher energy (see Section~\ref{sec:flareshape}).

For the simple, classical flare, we draw on the \citet{Davenport2014} empirical flare template, which parameterizes the flare as a fast rise phase
\begin{equation} \frac{F_{\rm rise}}{F_{\rm amp}} = 1 + 1.941 t_{1/2} - 0.175 t_{1/2}^2 - 2.246 t_{1/2}^3 - 1.125 t_{1/2}^4 \end{equation}
and a two-component exponential decay phase
\begin{equation} \frac{F_{\rm decay}}{F_{\rm amp}} = 0.689 e^{-1.600 t_{1/2}} + 0.303 e^{-0.278 t_{1/2}}. \end{equation} 
The flare is entirely characterized by the flare amplitude ($F_{\rm amp}$) and the half-light decay timescale ($t_{\rm 1/2}$).  \citet{Davenport2014} created this flare template from the co-added lightcurves of the 885 simple, classical flares found in \textit{Kepler} data for M4 GJ~1243, the most active M dwarf in the original \textit{Kepler} field. In addition to $F_{\rm amp}$ and $t_{\rm 1/2}$, the agreement between the flare template and the flare observations depends sensitively on the position of the template peak relative to the integration times of the observed data point. Because we are calculating the lower limit on the $V$-band flare energy, we assume that the actual flare peak occurred during the ASAS-SN observations, and let the exact position of the template peak vary between the times of the first and last observed data points.

To estimate the best fit of the flare template to the observations, we first generated 400 flare light curves with $t_{\rm 1/2}$ varying from 10 to 2000~s in increments of 5~s. We integrated the flare lightcurve over the exposure time for each data point (90~s), varying the position of the peak from the beginning of the first exposure to the end of the last exposure in increments of 1~s. For each combination of peak position and $t_{\rm 1/2}$ we compared the integrated fluxes to the data points, accepting the parameters where the normalized integrated light-curve fell within the 1-$\sigma$ uncertainties of the observed fluxes. 

\begin{figure*}
\includegraphics[width=0.95\linewidth]{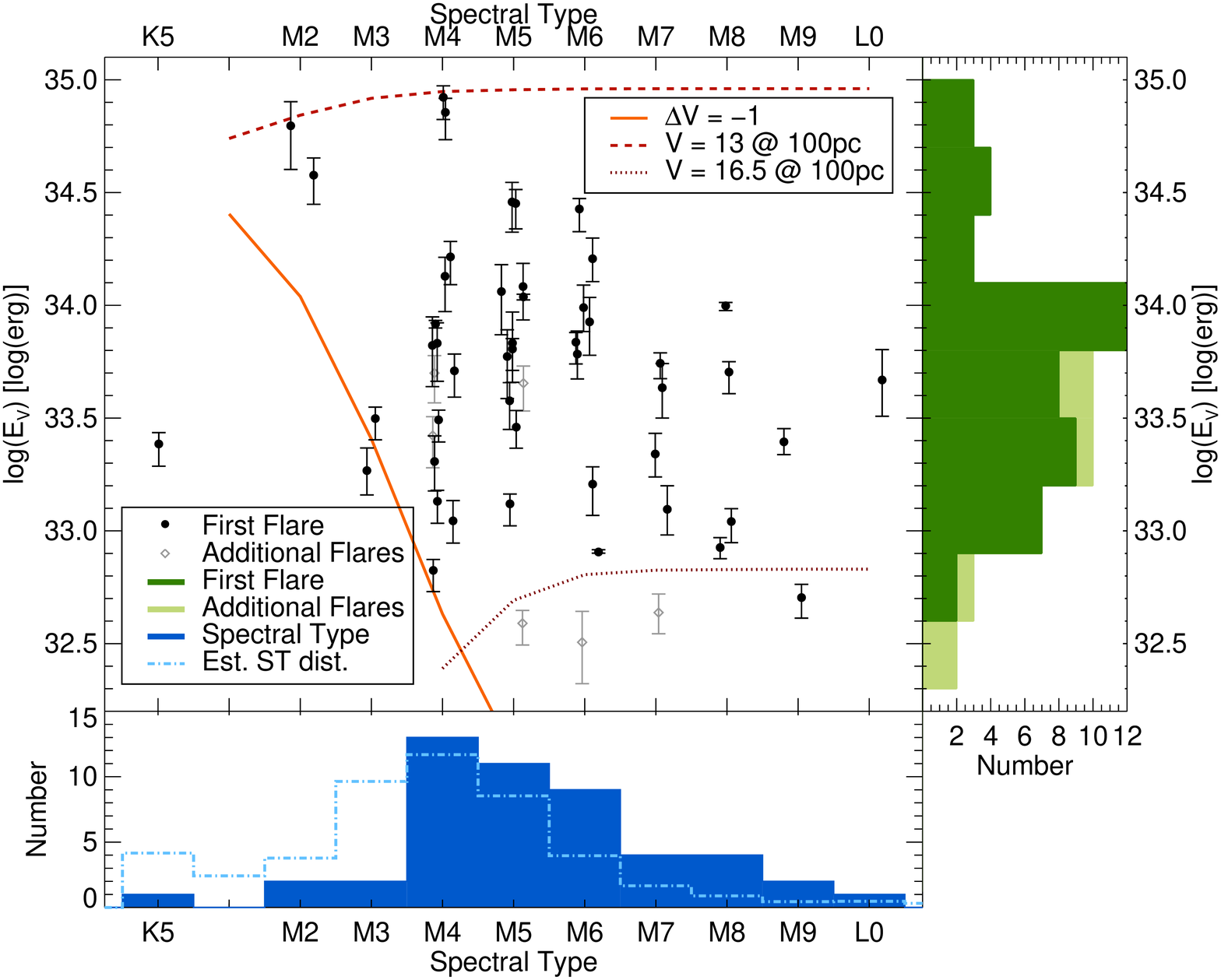} 
\caption{The log of the lower limit on $V$-band energy ($\log(E_V)$) as a function of spectral type. The lower energy flares from M dwarfs with multiple flares (grey points; light green histogram) are distinguished from the main sample of flares (black points, dark green histogram). The lower limit $V$-band energies range from $\log(E_V/{\rm ergs}) = 32$ to $35$, and are set by the detection limit of ASAS-SN ($\sim V<16.5$ mag, as shown in the red dashed lines). The spectral type distribution (dark blue) is comparable to a distribution drawn from the \citet{Bochanski2011} M dwarf luminosity function (light blue dot dashed line) where M3 and earlier dwarfs are under-represented due to the ASAS-SN detection limits. The lack of flares in the bottom left corner is an observational effect; these flares would likely be smaller than $\Delta V=-1.0$ (as shown by the orange line) and so unlikely to be found in the ASAS-SN search.} \label{fig:st_nrg}
\end{figure*}

The parameter space where the flare template and observations are in agreement is not distributed normally with respect to $t_{\rm 1/2}$ or peak position, so we adopt the median values as the best fit result and the interquartile range as the asymmetric uncertainties. We obtain the $V$-band energy released from the best-fit flare template by integrating over total flare time, and multiplying by the distance factor ($d^2$) and $V$-band central wavelength (5500~\AA). The resulting $V$-band energy is then a lower limit of the observed ASAS-SN flare. These lower limits on the $V$-band energy range from $\log(E_V/{\rm ergs})=32$ to $35$ are shown as a function of spectral type in Figure~\ref{fig:st_nrg} and listed in Table~\ref{tab:flares}. 

We examined the observations and the range of best-fit template curves for each flare to classify each set of data points as occurring during the rise or decay phase of each flare. We found that of the 54 flares with more than one photometric observation, we observed the rise phase 16 times, the decay phase 37 times, and we caught the peak in the middle of three observations for one flare. Over the course of an entire flare, the decay phase is roughly 12 times longer than the rise phase \citep{Davenport2014}. It is unlikely that we are observing the end of the decay phase, however, since flares are much fainter then. The ratio of decay phase to rise phase observed for the ASAS-SN sample (2.31) corresponds to the ratio of decay phase to rise phase for the flare template if it is only visible at 52\% or more of its peak value. Hence, the balance of observations occurring during rise versus decay is a good indicator that we are likely observing most flares close to their peak values. 

\renewcommand{\arraystretch}{0.9}
\begin{deluxetable*}{l lll ll | llll | ll} \tablewidth{0pt} \tabletypesize{\scriptsize}
\tablecaption{Flare Properties \label{tab:flares} }
\tablehead{ \colhead{ASAS-SN} & \colhead{quiescent} & \colhead{flare}    & \colhead{$\Delta V$} & \colhead{flare} & \colhead{\#   } & \multicolumn{4}{c}{fit}                                                       & \multicolumn{2}{c}{from $t_{\rm half}$ relation} \\
            \colhead{number}  & \colhead{$V$}       & \colhead{peak $V$} & \colhead{}           & \colhead{type}  & \colhead{flare} & \colhead{$t_{\rm half}$} & \multicolumn{3}{c}{$log(E_V)$ [log(erg)]}          & \colhead{$t_{\rm half}$} & \colhead{$log(E_V)$}   \\
            \colhead{}        & \colhead{}          & \colhead{}         & \colhead{}           & \colhead{}      & \colhead{obs  } & \colhead{[s]}            & \colhead{IQ 1} & \colhead{median} & \colhead{IQ 3} & \colhead{[s]}            & \colhead{[log(erg)]} } 
\startdata
13ba & 17.42$\pm$0.01 & 12.04$\pm$0.01 & $-$5.38$\pm$0.01 & decay & 2 & 309 & 34.33 & 34.43 & 34.47 & 1927 & 35.14 \\
13be & 18.40$\pm$0.01 & 16.02$\pm$0.06 & $-$2.38$\pm$0.06 & decay & 2 & 509 & 32.95 & 33.04 & 33.13 & 573 & 33.09 \\
13bf & 17.22$\pm$0.01 & 14.33$\pm$0.03 & $-$2.89$\pm$0.03 & rise & 2 & 443 & 34.45 & 34.58 & 34.65 & 1942 & 35.15 \\
13bg & 17.13$\pm$0.01 & 14.82$\pm$0.04 & $-$2.31$\pm$0.04 & rise & 4 & 1373 & 33.07 & 33.21 & 33.28 & 458 & 32.71 \\
13bh & 16.71$\pm$0.05 & 13.94$\pm$0.02 & $-$2.76$\pm$0.05 & decay & 2 & 282 & 33.71 & 33.81 & 33.85 & 1217 & 34.36 \\
13bi & 18.08$\pm$0.01 & 14.81$\pm$0.03 & $-$3.27$\pm$0.04 & rise & 2 & 1261 & 33.66 & 33.83 & 33.92 & 760 & 33.57 \\
13bk & 18.68$\pm$0.01 & 15.11$\pm$0.07 & $-$3.57$\pm$0.07 & single & 1 & \nodata & \nodata & \nodata & \nodata & 1008 & 34.04 \\
13bl & 14.83$\pm$0.01 & 13.17$\pm$0.01 & $-$1.66$\pm$0.02 & decay & 2 & 972 & 33.40 & 33.50 & 33.55 & 650 & 33.30 \\
13bn & 17.16$\pm$0.02 & 15.08$\pm$0.06 & $-$2.08$\pm$0.06 & rise & 2 & 1315 & 33.97 & 34.13 & 34.21 & 953 & 33.95 \\
13bt & 19.14$\pm$0.06 & 14.62$\pm$0.03 & $-$4.52$\pm$0.07 & rise & 2 & 236 & 33.50 & 33.63 & 33.74 & 1132 & 34.24 \\
13cb & 21.97$\pm$0.15 & 12.85$\pm$0.01 & $-$9.13$\pm$0.15 & decay & 7 & 270 & 33.61 & 33.70 & 33.75 & 1137 & 34.25 \\
13cm & 16.21$\pm$0.03 & 13.43$\pm$0.02 & $-$2.78$\pm$0.04 & decay & 4 & 545 & 33.39 & 33.49 & 33.54 & 777 & 33.60 \\
13cr & 14.82$\pm$0.02 & 12.41$\pm$0.01 & $-$2.41$\pm$0.02 & rise & 2 & 1471 & 34.73 & 34.86 & 34.92 & 1640 & 34.86 \\
13de & 18.14$\pm$0.02 & 15.38$\pm$0.05 & $-$2.76$\pm$0.06 & rise & 2 & 877 & 33.59 & 33.71 & 33.78 & 799 & 33.65 \\
13di & 19.97$\pm$0.06 & 16.00$\pm$0.11 & $-$3.97$\pm$0.12 & rise & 2 & 1283 & 34.32 & 34.46 & 34.54 & 1262 & 34.42 \\
13dj & 23.09$\pm$0.04 & 17.35$\pm$0.23 & $-$5.74$\pm$0.24 & decay & 2 & 1188 & 34.60 & 34.80 & 34.90 & 1702 & 34.93 \\
13dk & 19.91$\pm$0.02 & 15.81$\pm$0.11 & $-$4.10$\pm$0.11 & rise & 2 & 562 & 33.45 & 33.58 & 33.66 & 833 & 33.72 \\
14bj & 19.73$\pm$0.01 & 16.71$\pm$0.13 & $-$3.02$\pm$0.13 & rise & 2 & 838 & 33.18 & 33.31 & 33.42 & 608 & 33.19 \\
14bm & 20.47$\pm$0.01 & 16.10$\pm$0.09 & $-$4.37$\pm$0.09 & decay & 2 & 1245 & 33.94 & 34.08 & 34.19 & 953 & 33.95 \\
14bn & 17.40$\pm$0.04 & 14.46$\pm$0.04 & $-$2.94$\pm$0.05 & decay & 2 & 290 & 33.03 & 33.13 & 33.18 & 711 & 33.45 \\
14cx & 19.62$\pm$0.02 & 14.33$\pm$0.03 & $-$5.29$\pm$0.04 & decay & 2 & 1254 & 34.11 & 34.21 & 34.30 & 1065 & 34.13 \\
14dv & 20.72$\pm$0.01 & 14.67$\pm$0.06 & $-$6.05$\pm$0.06 & decay & 2 & 614 & 33.88 & 33.99 & 34.09 & 1143 & 34.25 \\
14ea & 15.78$\pm$0.01 & 13.77$\pm$0.02 & $-$2.01$\pm$0.02 & decay & 2 & 360 & 33.02 & 33.12 & 33.16 & 658 & 33.32 \\
        &              & 14.69$\pm$0.04 & $-$1.09$\pm$0.05 & decay & 2 & 336 & 32.49 & 32.59 & 32.65 & 447 & 32.67 \\
14fi & 17.21$\pm$0.03 & 13.58$\pm$0.03 & $-$3.63$\pm$0.04 & rise & 2 & 1317 & 34.34 & 34.45 & 34.51 & 1245 & 34.40 \\
14gj & 20.63$\pm$0.05 & 15.68$\pm$0.06 & $-$4.95$\pm$0.08 & decay & 2 & 355 & 33.37 & 33.46 & 33.53 & 882 & 33.82 \\
14gn & 21.38$\pm$0.04 & 15.80$\pm$0.25 & $-$5.58$\pm$0.25 & rise & 2 & 1061 & 33.66 & 33.83 & 33.97 & 871 & 33.79 \\
        &              & 15.88$\pm$0.08 & $-$5.50$\pm$0.09 & rise & 2 & 637 & 33.53 & 33.65 & 33.73 & 848 & 33.75 \\
14hc & 18.99$\pm$0.02 & 14.10$\pm$0.04 & $-$4.89$\pm$0.04 & decay & 2 & 181 & 33.29 & 33.39 & 33.44 & 993 & 34.02 \\
14hz & 20.16$\pm$0.02 & 15.42$\pm$0.06 & $-$4.74$\pm$0.06 & decay & 2 & 699 & 33.67 & 33.78 & 33.89 & 932 & 33.91 \\
14ji & 22.42$\pm$0.05 & 16.92$\pm$0.23 & $-$5.50$\pm$0.23 & rise & 3 & 1260 & 33.24 & 33.34 & 33.43 & 559 & 33.05 \\
        &              & 16.19$\pm$0.13 & $-$6.23$\pm$0.14 & decay & 2 & 76 & 32.54 & 32.64 & 32.72 & 703 & 33.43 \\
14jw & 18.96$\pm$0.06 & 14.54$\pm$0.04 & $-$4.42$\pm$0.07 & decay & 2 & 981 & 32.98 & 33.10 & 33.20 & 485 & 32.81 \\
14jy & 15.05$\pm$0.01 & 12.10$\pm$0.01 & $-$2.95$\pm$0.01 & decay & 2 & 1421 & 34.09 & 34.21 & 34.28 & 1005 & 34.04 \\
        &              & 12.05$\pm$0.01 & $-$3.00$\pm$0.01 & rise & 2 & 379 & 33.57 & 33.70 & 33.78 & 1020 & 34.06 \\
        &              & 13.72$\pm$0.02 & $-$1.33$\pm$0.02 & rise & 4 & 1321 & 33.28 & 33.42 & 33.51 & 548 & 33.02 \\
14ke & 21.03$\pm$0.04 & 15.29$\pm$0.06 & $-$5.74$\pm$0.07 & decay & 2 & 143 & 32.95 & 33.04 & 33.10 & 811 & 33.67 \\
14lc & 24.20$\pm$0.10 & 14.86$\pm$0.07 & $-$9.34$\pm$0.12 & decay & 2 & 97 & 32.61 & 32.70 & 32.76 & 686 & 33.39 \\
14mz & 19.38$\pm$0.02 & 13.17$\pm$0.01 & $-$6.21$\pm$0.02 & decay & 2 & 548 & 33.74 & 33.84 & 33.88 & 1016 & 34.05 \\
        &              & 16.89$\pm$0.28 & $-$2.49$\pm$0.28 & middle & 3 & 1078 & 32.32 & 32.51 & 32.64 & 303 & 32.03 \\
15ep & 19.66$\pm$0.03 & 15.68$\pm$0.16 & $-$3.98$\pm$0.16 & decay & 2 & 1177 & 33.59 & 33.77 & 33.89 & 767 & 33.58 \\
15kl & 17.21$\pm$0.04 & 14.64$\pm$0.05 & $-$2.57$\pm$0.06 & decay & 2 & 909 & 33.16 & 33.27 & 33.37 & 569 & 33.08 \\
15ll & 22.60$\pm$0.03 & 14.89$\pm$0.04 & $-$7.71$\pm$0.05 & decay & 2 & 151 & 34.82 & 34.92 & 34.97 & 3493 & 36.15 \\
15oy & 18.63$\pm$0.01 & 13.92$\pm$0.04 & $-$4.71$\pm$0.04 & decay & 3 & 1401 & 32.88 & 32.93 & 32.97 & 378 & 32.39 \\
15tv & 19.66$\pm$0.01 & 13.61$\pm$0.02 & $-$6.05$\pm$0.02 & decay & 3 & 341 & 34.02 & 34.04 & 34.05 & 1415 & 34.61 \\
16ae & 23.75$\pm$0.34 & 13.51$\pm$0.02 & $-$10.24$\pm$0.34 & decay & 3 & 595 & 33.51 & 33.67 & 33.80 & 942 & 33.93 \\
16cx & 22.81$\pm$0.18 & 15.01$\pm$0.04 & $-$7.80$\pm$0.19 & decay & 3 & 1082 & 33.34 & 33.39 & 33.45 & 597 & 33.16 \\
16di & 20.36$\pm$0.01 & 15.79$\pm$0.07 & $-$4.57$\pm$0.07 & decay & 6 & 1533 & 33.67 & 33.74 & 33.79 & 672 & 33.36 \\
16dj & 16.03$\pm$0.01 & 12.29$\pm$0.01 & $-$3.74$\pm$0.01 & decay & 3 & 668 & 33.90 & 33.92 & 33.93 & 1045 & 34.10 \\
16dr & 19.60$\pm$0.02 & 13.87$\pm$0.02 & $-$5.73$\pm$0.03 & decay & 3 & 1157 & 33.87 & 34.06 & 34.18 & 996 & 34.02 \\
16du & 21.80$\pm$0.07 & 16.17$\pm$0.14 & $-$5.63$\pm$0.16 & decay & 3 & 1222 & 33.78 & 33.93 & 34.03 & 871 & 33.79 \\
16gt & 17.09$\pm$0.02 & 12.69$\pm$0.01 & $-$4.40$\pm$0.02 & decay & 3 & 253 & 32.90 & 32.91 & 32.92 & 609 & 33.19 \\
16hq & 21.55$\pm$0.02 & 13.35$\pm$0.02 & $-$8.20$\pm$0.03 & decay & 3 & 469 & 33.98 & 34.00 & 34.01 & 1242 & 34.39 \\
16kq & 18.17$\pm$0.01 & 14.56$\pm$0.04 & $-$3.61$\pm$0.04 & decay & 3 & 1096 & 33.64 & 33.82 & 33.95 & 829 & 33.71 \\
GJ3039      & 12.66$\pm$0.01 & 11.66$\pm$0.01 & $-$1.00$\pm$0.01 & decay & 2 & 1142 & 32.73 & 32.82 & 32.87 & 362 & 32.33 \\
\enddata
\end{deluxetable*}
\renewcommand{\arraystretch}{1.0}

\subsection{Flare Timescales}
\label{sec:ftscl}
\begin{figure}
\includegraphics[width=0.95\linewidth]{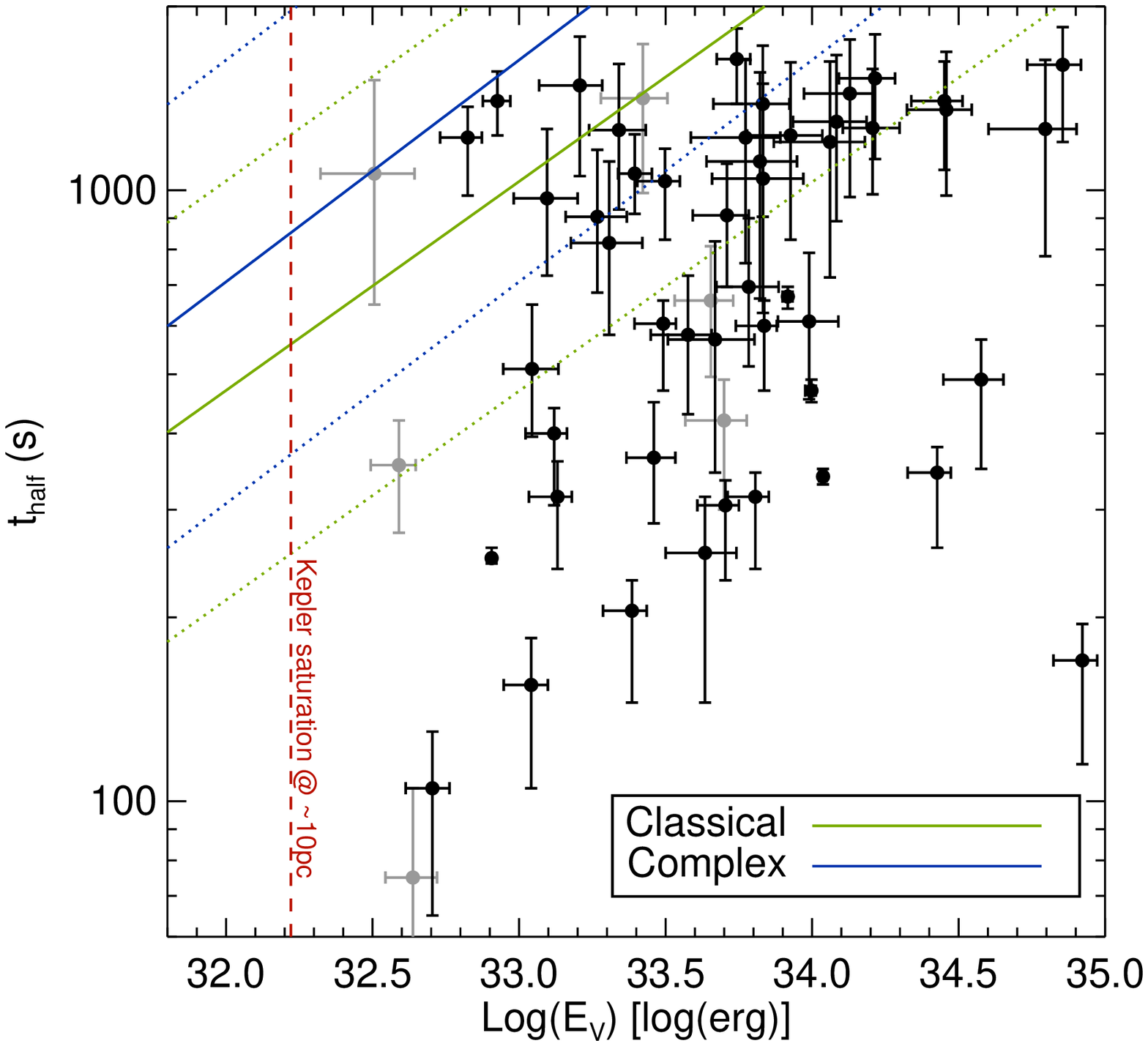} 
\caption{Flare timescale ($t_{1/2}$) as a function of the log of the best fit $V$-band energy. The ASAS-SN M-dwarf flares are compared to estimates of flare timescale based on the correlation between total flare duration and \textit{Kepler} flare energy from \citet{Hawley2014} and \citet{Silverberg2016}, shown for both classical (green) and complex (blue) flares. For each relationship, an order of magnitude spread in energy is shown (dotted lines) to demonstrate the range observed in flare data. A characteristic saturation limit for \textit{Kepler} observations of flares on M dwarfs at 10pc is shown, based on the saturated data from GJ 1243 (red dashed line). The majority of \textit{Kepler} flares analyzed in detail have energies below this line, so the relationships shown are extrapolations for energies greater than $\log(E_V/{\rm ergs}) > 32.2$.} \label{fig:tscl}
\end{figure}

The timescales for each flare are determined from one or two points. To test whether the resulting flare timescales are reasonable, we compare to the correlations between flare energy and flare length found in high cadence M dwarf observations from \textit{Kepler} data \citep{Hawley2014,Silverberg2016}. They derive two linear relationships, one for classical flares and one for multi-peaked complex flares. We convert the relationship between flare duration and \textit{Kepler}-band energy to $t_{1/2}$ and $V$-band energy assuming $t_{1/2} = 0.14\times t_{\rm duration}$ \citep{Davenport2016} and a flare SED of a 10,000~K blackbody \citep[see, e.g.,][]{Kowalski2013} to convert energies between the two bands ($E_{Kepler} = 1.9E_V$). Figure~\ref{fig:tscl} shows the relationship between timescale and flare energy compared to the timescales and energies derived from the fitting procedure described in the previous section. We also calculate energies for each flare based on that relationship; because flare energy and $t_{1/2}$ are dependent on each other, we calculate them iteratively until the change in the log of energy is less than 0.01 between steps. The final values (given in Table~\ref{tab:flares}) fall within an order of magnitude of those based on the fitting procedure, and are on average a few percent stronger. 

Compared to the relationship from \textit{Kepler} data, approximately half of the flares have timescales and energies that are in the expected range based on the \textit{Kepler} flare data, but the remaining half have shorter timescales than expected for flares with their peak flux. The majority of the ASAS-SN flares are higher energy than those observed in \textit{Kepler} data of GJ~1243, however, so the relationship between duration and energy has been extrapolated to higher energies than it may be applicable. In a sample of $\sim20$ well-characterized flares, \citet{Kowalski2013} found that larger flares were more impulsive (shorter $t_{1/2}$) than smaller flares, indicating the relationship between duration and energy may be shallower at higher flare energies. The relationship between half-light timescale ($t_{1/2}$) and total duration is also based on a single flare peak, and so does not hold for complex flares that have multiple peaks. Complex flares would have shorter measured $t_{1/2}$ than the expected values for simple flares. Each of these factors lead to an underestimate of the total flare energy, so the $V$-band energies derived from the light-curve fits are conservative as lower limits. 

\subsection{Flares With Extra Observations}
\label{sec:multf}
\begin{figure}
\includegraphics[width=0.95\linewidth]{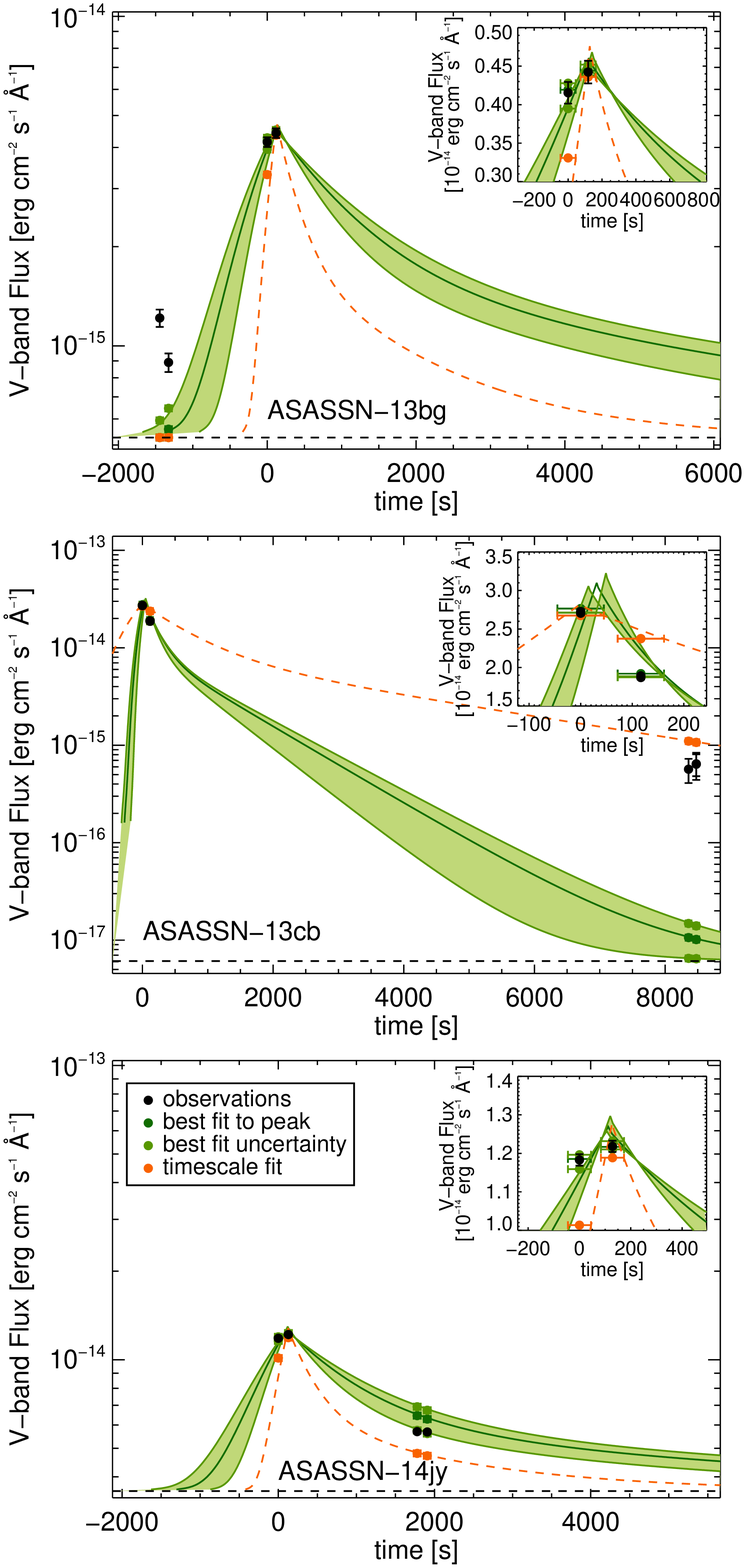} 
\caption{$V$-band flare flux from ASASSN-13bg (top), ASASSN-13cb (middle), and ASASSN-14jy (bottom) as a function of time, with the inset showing the same data near the peak of the flare. The flare observations are shown (black points with uncertainties) as well as the quiescent flux value (black dashed line). The best-fit flare template is shown (dark green line and points) with its uncertainties (light green shaded area), as well as a flare template based on the relationship between total energy and timescale from \textit{Kepler} data (orange line and points; see Section~\ref{sec:ftscl}). The data points generated from these each of these light-curves integrated over a 90~s exposure time are also shown. For 13cb and 14jy, the data points far from the peak do not fall near the best-fit curve, but for 14jy they do.} \label{fig:ep}
\end{figure}

Three of the ASAS-SN flare M dwarfs had additional observations from the same flare separated by more than a few minutes, and we examine these to better understand the limitations of our fit. For ASASAN-13bg and ASASSN-14jy, this occurred because the stars are positioned on the overlap strip between two different ASAS-SN fields, resulting in two sets of observations separated by $\sim15$ minutes. For ASASSN-13cb, the additional observations were an effort to obtain follow-up data for this dramatic transient event \citep[see][]{Schmidt2014}. During our fitting procedure, we excluded points farther than 1000~s from the peak to focus on the data from a singe telescope pointing, so the fits for those three objects are based on a subset of the available data. The light curves and fits for these objects are shown in Figure~\ref{fig:ep}.

The best fit to ASASSN-13bg is based on two observations at the end of the rise phase of the flare, and there are two additional observations taken before the peak of the flare that fall above the model. To fit the later observations would require a longer $t_{\rm 1/2}$ than the model shown, but the current best fit parameters already have a long half-light timescale ($t_{\rm 1/2}=1373$~s) compared to the $V$-band energy ($\log(E_V/{\rm ergs})=32.7$). Because they fall so far from the best fit to the peak observations, it is more likely that the points excluded from the fit are part of a different peak in the same flare. While the precursor peak could be very strong, a minimum fit to it would also be nearly an order of magnitude less energetic than the main flare and only contribute an additional few percent to the total energy. 

The best fit to ASASSN-13cb is based only on the two observations near the peak, not including two additional observations taken over two hours later. This resulted in a relatively short half-light timescale ($t_{\rm 1/2}=270$~s) and a $V$-band energy ($\log(E_V/{\rm ergs})=32.7$) much lower than the original estimate from \citet{Schmidt2014}, which included all points but did not have an empirical model for the flare. The timescale calculated from the \textit{Kepler} relationship between energy and timescale ($t_{\rm 1/2}=2563$~s) is an excellent fit to the later data points, however, and corresponds to a much higher energy ($\log(E_V/{\rm ergs})=34.4$). That timescale is a poor fit for the steep drop between the first two data points. 

The match between the best-fit flare template and the observations is much better for ASASSN-14jy, with the points not included in the initial fit falling along the curve representing the lower quartile in uncertainty. This could indicate that the simple flare model is a good fit to at least some of the flares, but additional complexity could have occurred between the detected points. 

Given that none of the data points for any of these three flares fall significantly below the best-fit lines, it is reasonable to continue to assume that the fits represent solid lower limits on the V-band flare energies. 

\subsection{Upper Limits on Flare Energies}
\label{sec:ulenergy}
While sparse observations can place a strong lower limit on flare energies, the upper limits of the observed flares are less clear. We examine possible upper limits from two different perspectives.  First, we examine the variations in flare shape that would also fit the data in the context of previous work. Second, we use flare frequency distributions to estimate the probability of more luminous flares. 

\subsubsection{Constraints on the Flare Shape}
\label{sec:flareshape}
\begin{figure*}
\includegraphics[width=0.95\linewidth]{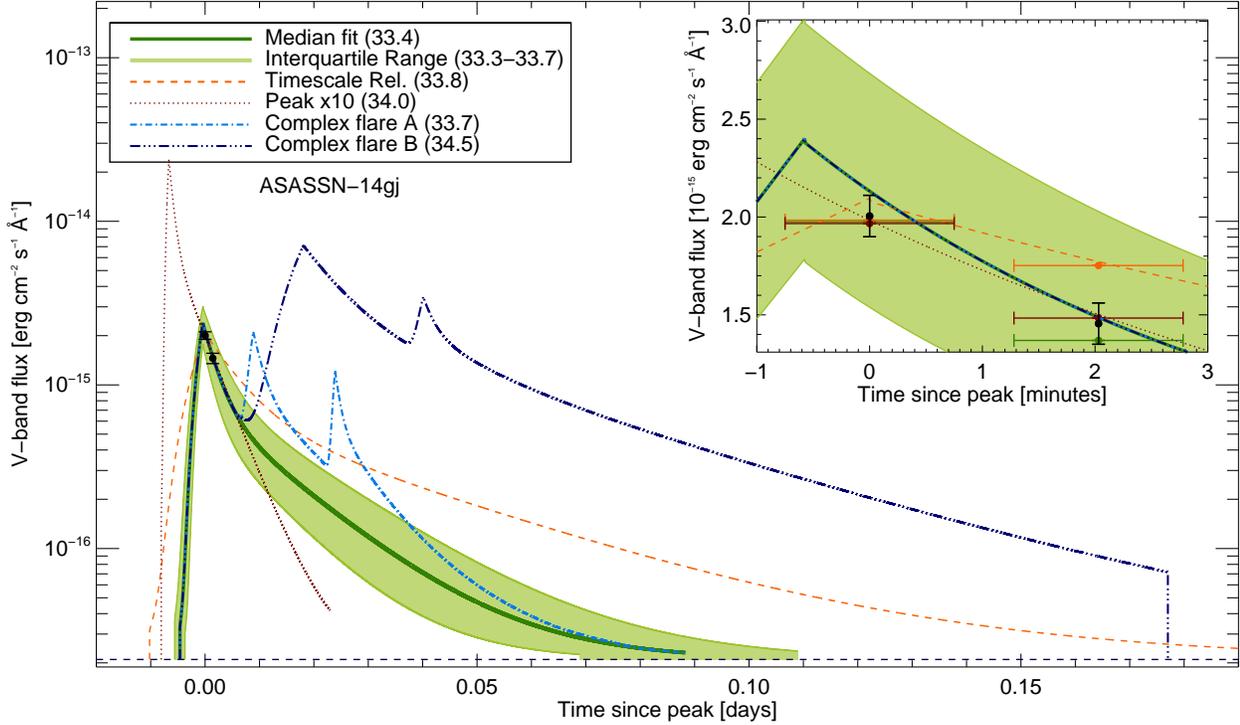} 
\caption{$V$-band flare flux from ASASSN-14gj as a function of time over 0.2~days, with an inset showing flare flux as a function of time over 4~minutes. The flare data is shown (black points with uncertainties) as well as the quiescent value (black dashed line). For each example flare shown, the full lightcurve is plotted (lines) along with the fluxes integrated over the exposure time of the data (90~s; points with horizontal error bars). The flare timescale fit to the data is shown (dark green line) with its uncertainties (light green shaded area), as well as a flare calculated to fall along the relationship between flare energy and duration found from \textit{Kepler} data for GJ~1243 \citep[orange dashed line;][]{Hawley2014,Silverberg2016}. A light curve with a peak at 10 times the observed peak is shown (red dotted lines), and two possible complex flares with multiple peaks are shown (dot dashed light blue and triple-dot dashed dark blue). For reference the log of the $V$-band energies under each of these curves is given in the legend.} \label{fig:lc}
\end{figure*}

In our energy estimates, we used the highest photometric observation as the peak of a classical flare template. We argue that these assumptions lead are both more likely to lead to a lower estimate of the flare energy, and so are valid lower limits. In Figure~\ref{fig:lc}, we use ASASSN-14gj as a reference to demonstrate some possible higher-energy templates. One of the simplest higher-energy flare templates is a classical shape with an earlier peak at a higher flux. We show a flare with a peak at 10 times the observed peak flux, with a decay timescale of 107~s and a total energy of $E_V=5.8\times10^{33}$~erg. The timescale is very short for this template due to the strong difference between the first two points, and would fall well below the relationship shown in Figure~\ref{fig:tscl}. The combination of timescale and energy is unlikely to be physically possible because the volume of hot plasma needed to produce the peak luminosity cannot cool that quickly. Thus, assuming a  classical flare template and the observed decay between dithered observations, it is likely that we discovered ASASSN-14gj near its peak. Because the timescales of the majority of ASAS-SN flares are already low compared to their energies, a similar argument can be made for the majority of the flares observed. 

While it seems likely that most of these flares were observed near their peak, the assumption that all these flares are simple, classical flares is less likely to be true. \citet{Davenport2014} find that approximately 15\% of the flares detected on GJ~1243 are complex, multi-peaked events. \citet{Hawley2014} and \citet{Silverberg2016} found that the fraction of complex flares increases with flare energy, with around half of flares above 10$^{33}$~ergs typically showing complex behavior. The higher incidence of complex events at high energies is, in part, because very energetic flares cover a relatively large area of the stellar surface and are more likely to trigger sympathetic flaring \citep[e.g.,][]{Anfinogentov2013}. Because the ASAS-SN flares are mostly higher energy than 10$^{33}$~erg, nearly half of the 55 flares are probably complex and so would be poorly fit by the empirical classical flare template.  Figure~\ref{fig:lc} shows two of many possibilities for complex, multi-peaked events, drawing inspiration from previously observed M dwarf flares \citep{Hawley1991,Kowalski2013}. It is not currently possible to estimate the presence or strength of multiple peaks based on sparse observations near the peak, but since the addition of more peaks increases the energy, the classical flare assumption continues to be a conservative lower limit. 

\subsubsection{The Frequency of Larger Flares}
\label{sec:ffdprob}
While the high occurrence rate of complex flares presents difficulties for placing an upper limit on a flare given a very sparse light curve, the distribution of flares at each energy is relatively well understood and can be used to place limits on flare energy. \citet{Lacy1976} first related the occurrence of flares to their energy, finding that on a handful of M-dwarfs, a flare that releases an order of magnitude more energy may occur ten to one-hundred times less frequently than smaller flares. This flare frequency distribution (FFD) has been observed and quantified for a number of stars, typically characterized by a power-law slope ranging from $-0.5$ to $-1.3$, dependent on spectral type and activity level. While our work eventually aims to quantify the FFD of M dwarfs, we can use previous results to examine the occurrence of large flares.

To place some upper limits on flare frequency, we use the upper and lower bounds of $\alpha$, the power law slope, from \citet{Hilton2011phd,Hawley2014}, to calculate a cumulative probability distribution of a given M-dwarf flare. If we assume no observational bias against flares of any size, each flare observed is increasingly less likely to be larger, proportional to their overall frequency. For example, the ASASSN-14gj flare with a lower limit of log(E) = 33.5 is 92\% likely to have an energy of less than 34.5, and 99\% likely to have an energy less than 35.5. Given the spread expected from the energy uncertainty and the range of possible $\alpha$ values, the flare is $>$70\% likely to be within an order of magnitude of the original estimate, and $>$97\% likely to be within two orders of magnitude.

\subsection{The Distribution of Flares and Stars}
The energies of flares and distribution of spectral types is essentially a scatter plot, with the shape of the distribution being set primarily by the underlying stellar population and the observational constraints rather than a fundamental property of flares. 

One limit on our detection of flares in early-M dwarfs is the contrast effect; on a dim, late-M dwarf, an energetic flare will have a much higher contrast with the underlying photosphere than it would on a bright, early-M dwarf. To show this effect, we calculated a characteristic energy for a one magnitude flare ($\Delta V = -1$) as a function of spectral type. To convert peak flux to energy, we used the relationship between flare peak and flare energy discussed in Section~\ref{sec:flareshape}. That limit (shown in Figure~\ref{fig:st_nrg}) falls below the majority of the flares for dwarfs earlier than M5. For dwarfs later than M5, flares smaller than $\log(E_V/{\rm ergs}) = 32$ would still have a contrast of greater than one magnitude. Due to the current classification methods of ASAS-SN, a flare of less than one magnitude is unlikely to have made it on our list. 

We also examined the effect of the faint detection limit of ASAS-SN on the range of energies detected. The detection limit varies from night to night, but is typically above $V$=16 (and can be as faint as $V$=17.5). At a distance of 100~pc, that limit corresponds to a flare energy of $\log(E_V/{\rm ergs}) = 33$, and there are fewer flares than expected that fall below that energy. Those flares are likely to have occurred during good conditions or on more nearby stars. 

To understand the distribution of spectral types, we calculated the distribution expected if the 49 M dwarfs were drawn evenly from the Solar neighborhood \citep[shown in Figure~\ref{fig:st_nrg}; based on the mass functions of][]{Cruz2007,Bochanski2010}. Overall, there are fewer M3 and earlier dwarfs than expected, and more later types. This is reasonable, given that flares of the same energy are more difficult to detect on these bright M dwarfs. 

\section{ASAS-SN Flare Stars as a Population}
\label{sec:stellarprop}
The majority of flare surveys are targeted at specific stars (often particularly active ones), with the goal of examining detailed light-curves or photometry. ASAS-SN has the advantage of an agnostic flare selection, detecting flares wherever they occur. We use kinematics and the presence and strength of H$\alpha$ emission to assess the ASAS-SN flare stars as a population to determine if these flares occurred on typical field stars or on peculiarly young and active stars. 

\subsection{Kinematics of the ASAS-SN M dwarfs}
Kinematics are often used as age indicators for populations of stars; young stars typically have cold orbits characterized by low velocities with respect to the Sun and positions near the Galactic plane, while older stars have warmer orbits and are more likely to have higher velocities and be located further from the Galactic plane.  The majority (45 of 49 objects) of ASAS-SN flare stars are located within 200~pc of the Galactic plane, with only two M dwarfs (ASASSN-13dj and ASASSN-15ll) found at Galactic heights greater than 400~pc (in this case, both below the plane). 

We adopted Gaia proper motions for the 42 stars with matches in the Gaia database. For five of the seven remaining stars (excluding ASASSN-13dj and ASASSN-15ll due to the lack of 2MASS detections) we calculated proper motions based on the difference between 2MASS and WISE coordinates. The Gaia and 2MASS/WISE proper motions are given in Table~\ref{tab:kine}). We combined these proper motions with distances to calculate tangential velocities ($V_{tan}$) for 47 dwarfs, which are also given in Table~\ref{tab:kine}. The overall mean tangential velocity is 29.8~km~s$^{-1}$, with a standard deviation of 22.6~km~s$^{-1}$. There is a single M dwarf with a velocity larger than 100~km~s$^{-1}$ (ASASSN-13be with $V_{tan}=117.5$~km~s$^{-1}$). If this outlier is removed, the mean tangential velocity is 27.9~km~s$^{-1}$, with a standard deviation of 18.7~km~s$^{-1}$. 

Comparing these to the data from \citet{Faherty2009} indicates that the M dwarfs in the ASAS-SN M-dwarf sample are slightly younger than M dwarfs in the Solar neighborhood, implying ages younger than $\sim$5~Gyr and likely membership in the thin disk population. But with a few dwarfs at high Galactic heights and fast $V_{tan}$, it is not likely that they are a universally young population, nor are they significantly younger than the Galactic disk. 

\subsection{Activity of the ASAS-SN M dwarfs}
As a manifestation of strong magnetic fields, the flare rate of M dwarfs is frequently correlated with other indicators of magnetic activity \citep{Kowalski2009,Hawley2014}. Our spectroscopic observations include the region surrounding the H$\alpha$ emission line, frequently used as an indicator of activity in these low-mass stars \citep{Gizis2000,West2008}. We measured H$\alpha$ equivalent widths (EW) using a custom IDL code that adopts the wavelength ranges and method described by \citet{West2011}. Of our 49 stars, only three (K dwarf ASASSN-14hc and two M dwarfs) had no detectable emission in the H$\alpha$ line. Both of the M dwarfs had spectra with sufficiently high signal-to-noise that an emission line with an EW of $\sim$0.75~\AA\ would be detectable; they either have H$\alpha$ in relatively weak emission, weak absorption, or an absence of the activity indicator. 

\begin{figure}
\includegraphics[trim={1.5cm 0 0 0}, width=1.0\linewidth]{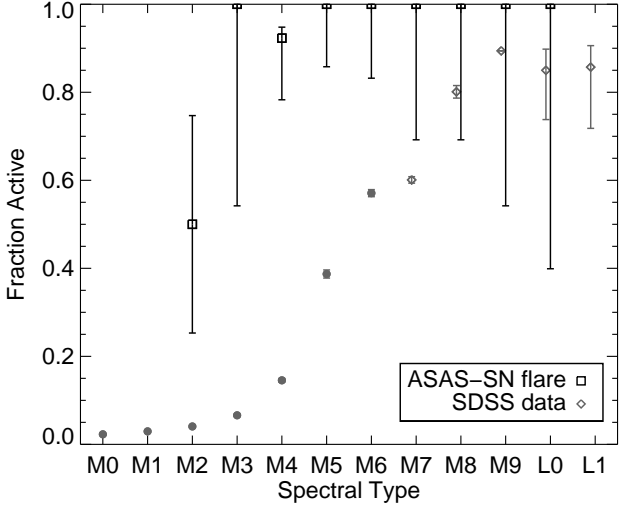} 
\caption{The fraction M dwarfs with activity (as defined by the presence of H$\alpha$ emission) for the ASAS-SN flare dwarfs (black) compared to a sample selected based on colors \citep[SDSS DR7 M dwarf and BUD ML dwarf samples; grey;][]{West2011,Schmidt2015}. The uncertainties are based on a binomial distribution. While the ASAS-SN flare sample is much smaller, it does also have a higher significantly active fraction than the comparison dwarfs.} \label{fig:frac}
\end{figure}

The activity fraction as a function of spectral type is shown in Figure~\ref{fig:frac} compared to the activity fraction of a reference sample of field M and L dwarfs that was selected based on color and not activity level \citep[the SDSS DR7 M dwarf sample and the BOSS Ultracool Dwarfs sample;][]{West2011,Schmidt2015}. The two M dwarfs without detected activity (13dj and 15ll) have spectral types of M2 and M4; at those spectral types, the mean activity fractions ($\sim5$\% and $\sim15$\% respectively) are relatively low, so the overall activity fraction of the ASAS-SN flare sample still falls well above the active fraction for those spectral types. These two dwarfs also have two of the strongest flares, and are the two most distant M dwarfs. Given their strong flares, it is surprising that they are not H$\alpha$ emitters, but it is possible they are slightly older M dwarfs that have a lower base level of activity and flare much more rarely. 

\begin{figure}
\includegraphics[trim={1.5cm 0 0 0}, width=1.0\linewidth]{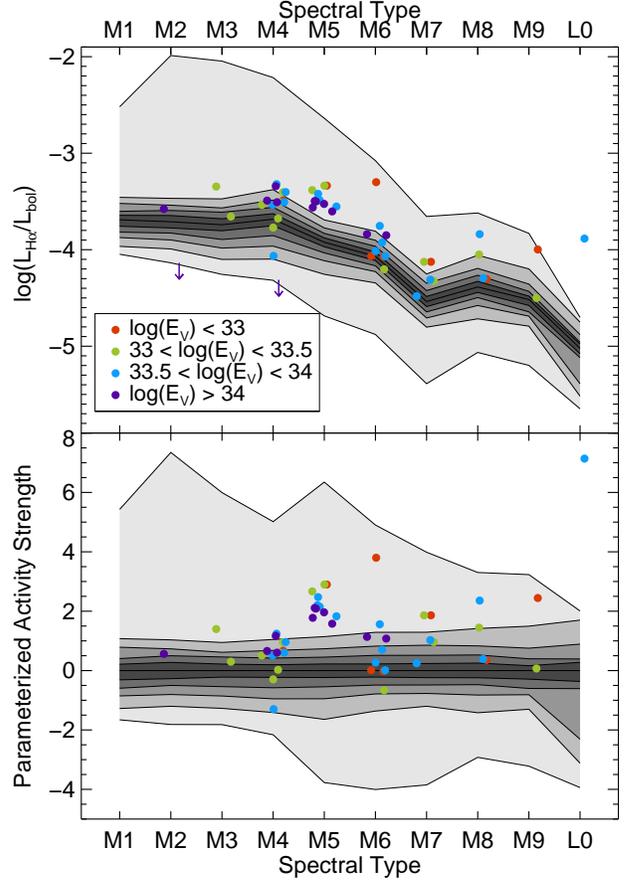} 
\caption{The log of the ratio of H$\alpha$ line luminosity to bolometric luminosity (H$\alpha$ activity strength) as a function of spectral type (top panel) and the parameterized activity strength (activity strength normalized so the median value is 0 and the interval between 30\% and 70\% is unity) as a function of spectral type (bottom panel) for the \citet{West2011} and \citet{Schmidt2015} M-dwarf samples compared to the activity strength and spectral type of the ASAS-SN flare sample. For the large M-dwarf samples, the shaded areas show percentiles in increments of 10\%. For the ASAS-SN flare sample, the different total flare energy is indicated by each point's color.  } \label{fig:halpha}
\end{figure}

The H$\alpha$ line strength can also be used to quantify the activity level of M dwarfs. EWs measure the strength of the line compared to the surrounding (psuedo-)continuum, but because the continuum varies with spectral type, we instead use the ratio of line luminosity to bolometric luminosity. We use the $\chi$ factor as a function of spectral type \citep{West2008a,Schmidt2014} to convert the EW to the ratio of H$\alpha$ emission line flux to the stellar bolometric luminosity. Figure~\ref{fig:halpha} shows the activity strength of the flaring ASAS-SN M dwarfs compared to that of the non-activity selected reference sample \citep{West2011,Schmidt2015}. We also show a parameterized activity strength, calculated by subtracting the median per spectral type from every value, then dividing by the difference between the 30\% and 70\% values. This parameterized activity strength is meant to aid comparisons of the ASAS-SN flare sample to the non-activity selected sample. 

The majority (42 our of 48) of the M and L dwarfs fall at activity levels above the median value for their spectral types. Those that have less than the median activity level include the two inactive dwarfs, as well as a few more M4 and one M6 dwarf. There are 27 M dwarfs with activity strength that falls above the 90th percentile values for their spectral type. This indicates that the ASAS-SN M dwarf flare sample is a significantly more active population than that of the field stars, consistent with previous results showing that stars without H$\alpha$ emission flare less often than those with \citep{Hawley2014}.

\section{Objects and Events of Interest}
\label{sec:indiv}
The majority of the objects in the ASAS-SN sample are newly discovered M dwarfs with typical colors and activity levels. Here we discuss M dwarfs with previous data and the peculiar objects observed as part of the ASAS-SN M-dwarf project. 

\subsection{GJ~3039}
\label{sec:GJ}
GJ~3039 (LP~525-39) was first identified as an H$\alpha$ emitting, high-proper motion source \citep{Stephenson1986a,Stephenson1986} before it was included in the third catalog of nearby stars \citep{Gliese1991} and assigned an M4 spectral type \citep{Reid1995}. \citet{McCarthy2001} found a companion to GJ~3039 with a separation of 0\farcs73 ($\sim30$~AU) and similar colors, but a magnitude difference of 0.7 in $J$-band. \citet{Schlieder2012a} identified GL~3039 as a candidate member of the $\beta$ Pictorus young moving group due to a combination of kinematics detected emission at X-ray in UV wavelengths. Previous photometric monitoring of GJ~3039 \citep{Messina2016} identified two small $R$-band flares. The ASAS-SN light curve of GJ~3039 also includes a few small ($\Delta V < -0.2$) flares, but none as large as the $\Delta V = -1.0$ flare examined here. 

\subsection{ASASSN-13cb (SDSS1022+19)}
\label{sec:13cb}
SDSS1022+19 was observed to flare in Fall 2013. Because the flare was both very large, and occurred on a very red object, follow-up observations were obtained rapidly and the object and its flare were analyzed by \citet{Schmidt2014}, who classified it as a field age M8 dwarf with a lower limit flare energy of log($E_V$) = 34.4 [log(erg)]. We re-analyze the flare here using an updated fitting procedure and the \citet{Davenport2014} flare template (unavailable during the original analysis) and find an energy of log($E_V$) = 33.7 [log(erg)]. As discussed in Section~\ref{fig:ep}, this fit does not show strong agreement with the late-stage ($\sim2$ hours later) decay observed for ASASSN-13cb, but does show agreement with the flare lightcurve calculated from the \textit{Kepler} relationship between timescale and energy. 

\subsection{ASASSN-15af (2M0518$-$05)}
\label{sec:tt}
\begin{figure*}
\includegraphics[width=0.95\linewidth]{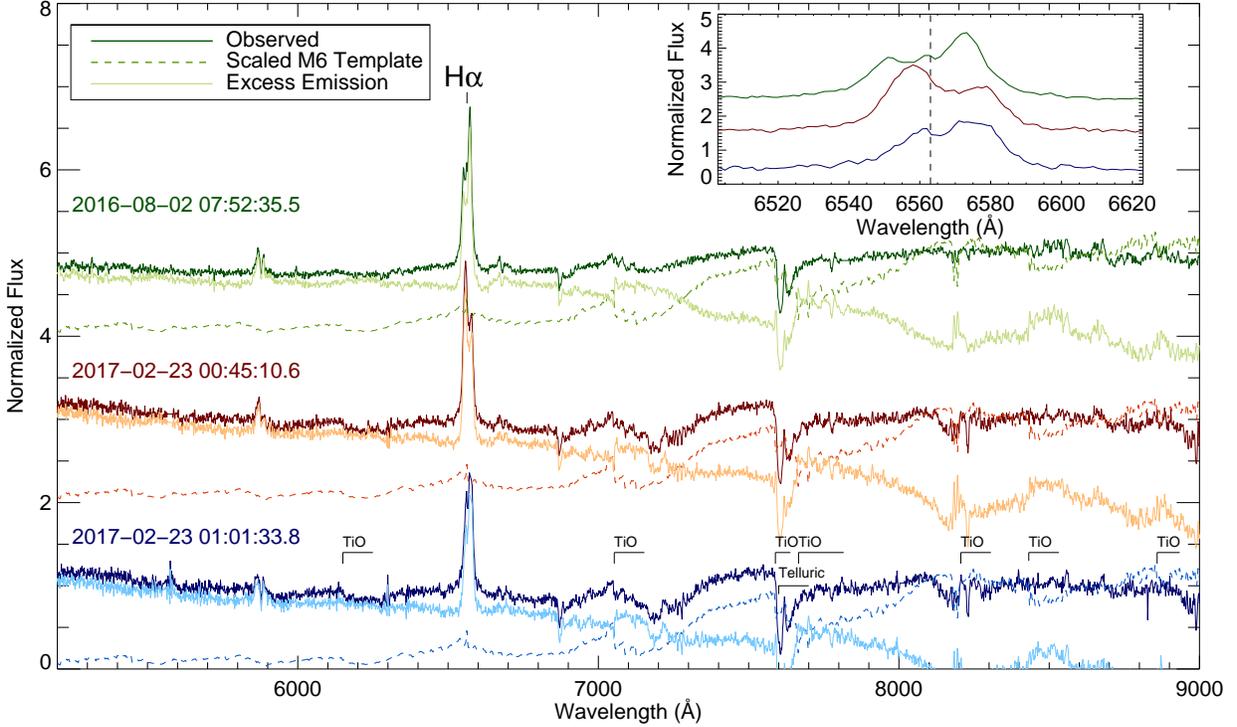} 
\caption{Three spectra of T Tauri star 2M0518$-$05 (dark green, red, and blue lines), observed on 2016 Aug 2 and 2017 Feb 23, compared to an M6 dwarf template (green, red and blue dashed lines). The T Tauri spectra with the M6 dwarf template subtracted (light green, red, and blue lines) show excess emission similar to that expected from flares, including elevated blue continuum and emission lines. The TiO features found in both the template and 2M0518$-$05 are labeled in the bottom spectrum. The H$\alpha$ line profiles (shown in the inset) are wide, variable, and multi-peaked, providing additional evidence that this object is showing activity primarily due to accretion rather than magnetic fields.} \label{fig:tt}
\end{figure*}

ASASSN-15af was identified as T Tauri star 2MASS~J05181685$-$0537300 (2M0518$-$05) as part of a photometric H$\alpha$ survey in Orion \citep{Wiramihardja1991}.  A spectrum from \citet{Lee2007} revealed many emission lines, including H$\alpha$ (EW$ = -57.3$\AA), OI 6300\AA, Fe 4924\AA, Ca H\&K, and He I (5876\AA), sufficient to classify 2M0518$-$05 as a T Tauri star, but there was no discussion of the continuum or spectral type. It is marked as a likely T Tauri star in the ASAS-SN transient page, but was placed on our observing list due to its red colors. We obtained three spectra of 2M0518$-$05 on 2016 August 2 and on 2017 Feb 23, shown in Figure~\ref{fig:tt}. 

The spectrum of 2M0518$-$05 is shown compared to an M6 template spectrum; while the blue continuum is not consistent with an M6 dwarf, there are molecular bands throughout the redder portion of the spectrum (7000--9000\AA) that are well-matched by the template spectrum. The subtracted spectrum shows less evidence of these bands, and appears to primarily be composed of blue continuum and emission lines. While the combination of emission lines and blue continuum would be expected in a magnetically-triggered flare, the line profiles are wider than those typically found during a flare event, most notably the H$\alpha$ emission, which is $\sim$50\AA\ wide and double peaked. It is more likely that the observed spectra are elevated due to accretion and other star formation processes. 

\begin{figure*}
\includegraphics[width=0.95\linewidth]{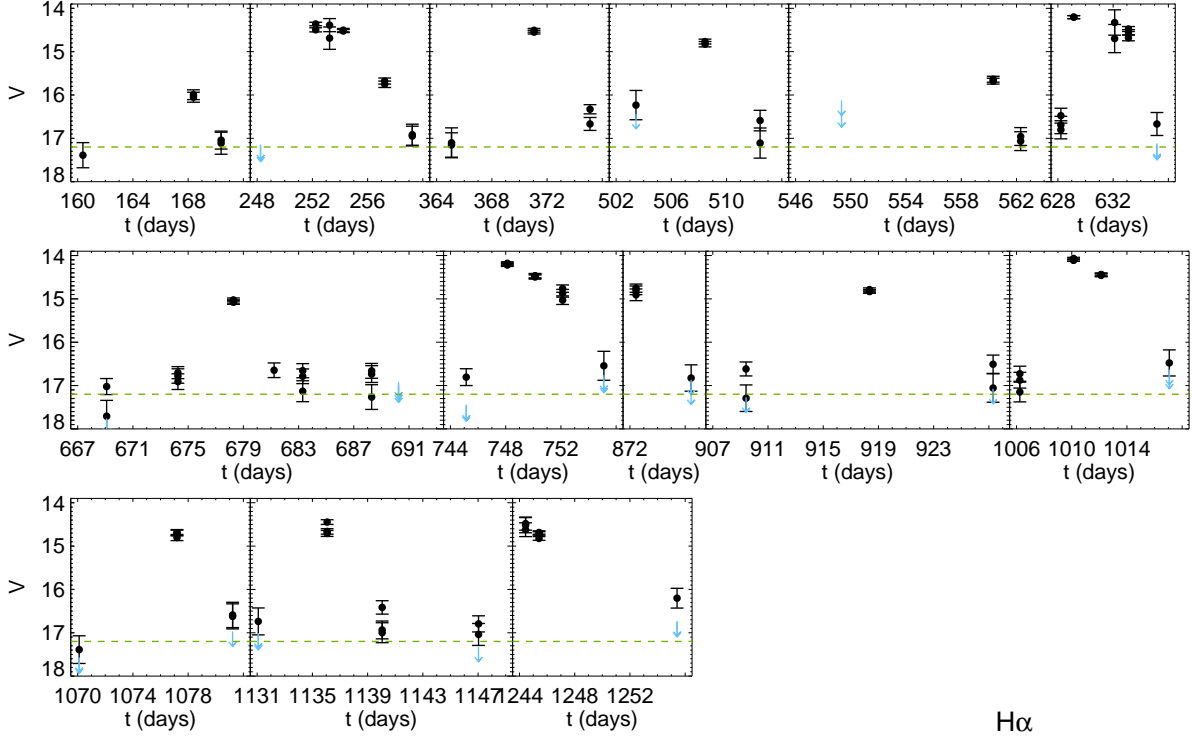} 
\caption{V magnitude as a function of time for 14 outbursts from 2M0518$-$05. A possible quiescent level ($V=17.2$) is shown (green dashed line) as well as the detections (filled circles) and upper limits (blue arrows). Time is given in days since 2014 Jan 1. Many outbursts may have lasted a similar time to a magnetically induced stellar flare (hours), but the light curve of 2M0518$-$05 also features a few multi-day outbursts that are most likely related to accretion events.} \label{fig:lc15af}
\end{figure*}

The ASAS-SN light curve of 2M0518$-$05 shows at least fourteen distinct outburst events, including the initial 2015 event that triggered the observations (all shown in Figure~\ref{fig:lc15af}). We obtained two additional photometric points for the object, obtaining $V=17.2$~mag both times. This is likely to be the quiescent value, as ASAS-SN detects 2M0518$-$05 in quiescence with a similar magnitude when the observing conditions are particularly good. While some of the outburst events were only detected on a single day and could be consistent with either magnetically induced flares or accretion, others last for multiple days at a time and so are far more likely to be accretion events. In addition to this activity, 2M0518$-$05 also shows strong variation in the CRTS database and has excess UV emission \citep{Sanchez2014}. Because any magnetic activity on this object is difficult to separate from accretion, we exclude 2M0518$-$05 from the sample of flaring M dwarfs. 
  
\subsection{ASASSN-16ae (SDSS0533+00)}
\label{sec:16ae}
SDSS0533+00 was observed to flare in January 2016. Because the flare was both very large, and occurred on a very red object, follow-up observations were obtained rapidly. The object and its flare were analyzed by \citet{Schmidt2016a}, who classified it as a field age L0 dwarf with a minimum $V$-band flare energy of $\log(E_V/{\rm ergs}) =  33.7$. The updated fitting method used in this work also measures $\log(E_V/{\rm ergs}) =  33.7$, though using a different distance; \citet{Schmidt2016a} used a combination of eight different photometric and spectrophotometric distances with a final result of d=95.5~pc, while here we adopt d=77.9~pc based on the $M_i$/$i-z$ relation from \citet[][in prep.]{Schmidt2018}.

\subsection{ASASSN-16hl}
\label{sec:mys}
\begin{figure}
\includegraphics[width=0.95\linewidth]{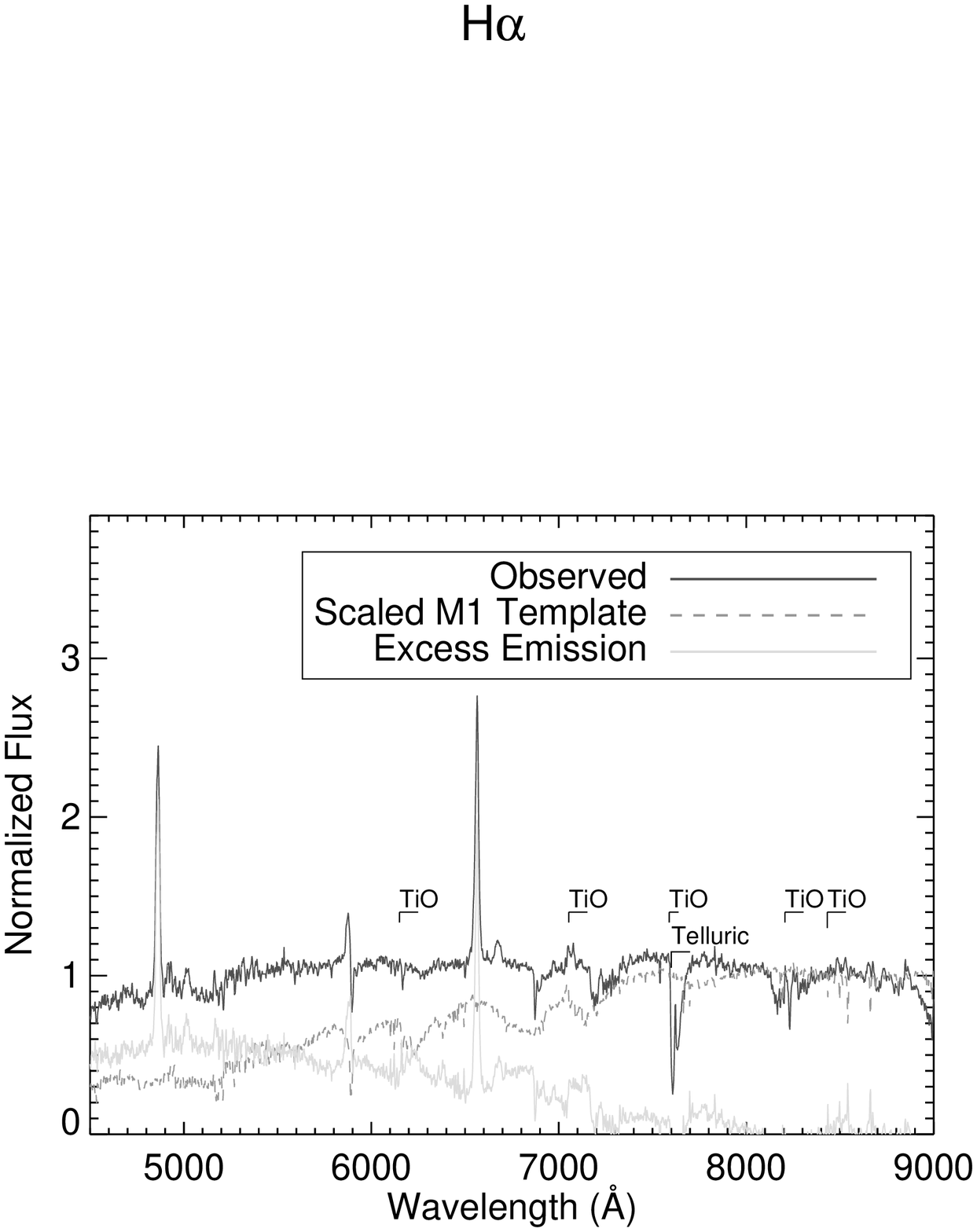} 
\caption{Spectrum of ASASSN-16hl, a red object with multiple outbursts in the ASAS-SN database. The object is compared to an M1 template and is possibly a late-type star with additional emission in the continuum. The TiO features found in both the template and ASASSN-16h are labeled. } \label{fig:16hl}
\end{figure}

ASASSN-16hl (2MASS J16045515-7223199) was marked in the ASAS-SN transients database as a CV or an outburst from a red star, and was selected for follow-up due to its particularly red colors in 2MASS and WISE. The spectrum of ASASSN-16hl (shown in Figure~\ref{fig:16hl}) has molecular absorption similar to an M1 dwarf, but has a blue continuum and emission lines similar to a flare or other hot outburst. No other spectra were obtained, so it is unclear whether the single spectrum represents a continuous or transient state. 

\begin{figure*}
\includegraphics[trim={0 8cm 0 0},width=0.95\linewidth]{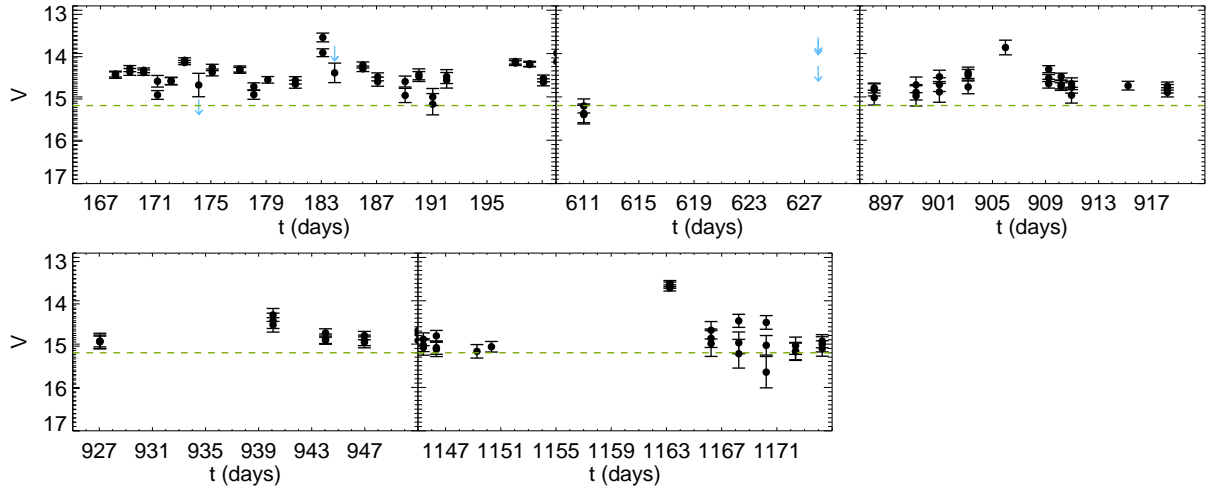} 
\caption{V magnitude as a function of time for three outbursts and two long timescale elevations of ASASSN-16hl.  } \label{fig:lc16hl}
\end{figure*}
The light curve of ASASSN-16hl shows multiple strong outbursts, including a strong event that resembles a flare (in part due to the lack of observations within a few days), two multi-day outbursts, and a few smaller, longer outbursts (shown in Figure~\ref{fig:lc16hl}). Our follow-up photometry yielded a magnitude of $V=14$, which is likely to have been during another outburst. The median ASAS-SN value is $V=16.1$~mag, which is possibly the quiescent value. In further support that this is a peculiar, young object, the WISE database flagged the photometry as variable, and a Simbad search reveals a possibly associated X-ray source in ROSAT within a few arcseconds. ASASSN-16hl is most likely a previously unclassified young star. We used the Banyan $\Sigma$ tool\footnote{\url{http://www.exoplanetes.umontreal.ca/banyan/banyansigma.php}} \citep{Gagne2018} to check if it is a member of any young moving group or association, and found it is most likely a field star. 

\section{Summary}
\label{sec:summary}
We selected 55 candidate flaring M dwarfs from ASAS-SN and performed follow-up spectroscopy and photometry on 53 of them, identifying four as peculiar objects, and 49 as K, M, or L dwarfs. From these 49 stars we identified 55 total flares. The flares have lower limits on their $V$-band energies that range from $\log(E_V/{\rm ergs})=32$ to 35, placing them among some of the strongest M-dwarf flares detected. The flare stars tend to have more and stronger chromospheric activity (as characterized by H$\alpha$ emission) than typical field M dwarfs, but are not an older population than field M dwarfs. 

If we adopt the \citet{Maehara2012} definition of super flares (energies greater than 10$^{33}$~erg), the majority of the stellar flares observed by ASAS-SN (44 of 55) fall above that cut in the $V$-band, and with \textit{Kepler} energies approximately 1.9 times larger than those in $V$-band, 50 of the 55 flares would fall into the super flare category. With its bright magnitude range and full sky coverage, ASAS-SN is well-suited to detecting these flares. The few largest ASAS-SN flares previously analyzed are larger than the biggest flares that have been examined in detail \citep[$\sim10^{34}$~erg;][]{Hawley1991,Kowalski2010}, but still smaller than the largest found on M dwarfs in the \citet{Davenport2016} flare catalog, which range up to $\log(E_{Kepler})=36$~[log(erg)].

ASAS-SN and similar sparse cadence surveys have the potential to uniquely observe an unbiased sample of nearby M dwarfs, detecting the largest flares and allowing us to characterize those flares as a population. As observations continue, we will continue to add to our understanding of M dwarf flares. This is essential not only to understand the relationship between flare activity and underlying stellar properties such as mass, rotation, and age, but also to estimate how frequently life on extrasolar planets may be threatened by the largest flares. 

In the last half of 2017 ASAS-SN deployed three additional units in Texas, South Africa, and Chile and by the end of 2018 we will deploy a sixth unit to China.  These new units both increase our cadence and decrease our vulnerability to weather.  With the 5-unit ASAS-SN we cover the entire visible sky every 20 hours, increasing the number of M-dwarf flares we recover dramatically. Using the data presented here combined with additional flares from the expanded survey, we plan to derive rates of the largest M dwarf flares from ASAS-SN in a follow-up work.

\acknowledgements{
We thank Nidia Morrell for her contributions to ASAS-SN observing. We thank the Las Cumbres Observatory and its staff for its continuing support of the ASAS-SN project. ASAS-SN is supported by the Gordon and Betty Moore Foundation through grant GBMF5490 to the Ohio State University and NSF grant AST-1515927. Development of ASAS-SN has been supported by NSF grant AST-0908816, the Mt. Cuba Astronomical Foundation, the Center for Cosmology and AstroParticle Physics at the Ohio State University, the Chinese Academy of Sciences South America Center for Astronomy (CASSACA), the Villum Foundation and George Skestos.

BJS, MRD, and SDJ were supported or partially supported by NASA through Hubble Fellowships grants HST-HF-51348.001,  HST-HF2-51373.001, and HST-HF2-51375.001-A, respectively, awarded by the Space Telescope Science Institute, which is operated by the Association of Universities for Research in Astronomy, Inc., for NASA, under contract NAS 5-26555.
KZS and CSK are supported by NSF grants AST-1515927 and AST-1515876.
Support for JLP is in part provided by FONDECYT through the grant 1151445
and by the Ministry of Economy, Development, and Tourism's Millennium
Science Initiative through grant IC120009, awarded to The Millennium
Institute of Astrophysics,
SD is supported by Project 11573003 supported by NSFC.

This research was made possible through the use of the AAVSO Photometric All-Sky
Survey (APASS), funded by the Robert Martin Ayers Sciences Fund.
This research has made use of data provided by Astrometry.net
\citep{Lang2010}.  This research has also made use of NASA's Astrophysics Data System Bibliographic Services and the SIMBAD database, operated at CDS, Strasbourg, France.

This work is based in part on observations obtained at the MDM Observatory, operated by Dartmouth College, Columbia University, Ohio State University, Ohio University, and the University of Michigan. 

This paper used data obtained with the MODS spectrographs
built with funding from NSF grant AST-9987045 and the NSF
Telescope System Instrumentation Program (TSIP), with additional
funds from the Ohio Board of Regents and the Ohio State University
Office of Research. 

This paper is based on data acquired using the LBT. The LBT
is an international collaboration among institutions in the United
States, Italy, and Germany. LBT Corporation partners are: The University
of Arizona on behalf of the Arizona university system; Istituto
Nazionale di Astrofisica, Italy; LBT Beteiligungsgesellschaft,
Germany, representing the Max-Planck Society, the Astrophysical
Institute Potsdam, and Heidelberg University; The Ohio State University,
and The Research Corporation, on behalf of The University
of Notre Dame, University of Minnesota and University of Virginia.

This publication makes use of data products from the Two Micron All Sky Survey, which is a joint project of the University of Massachusetts and the Infrared Processing and Analysis Center/California Institute of Technology, funded by the National Aeronautics and Space Administration and the National Science Foundation. This publication also makes use of data products from the Wide-field Infrared Survey Explorer, which is a joint project of the University of California, Los Angeles, and the Jet Propulsion Laboratory/California Institute of Technology, funded by the National Aeronautics and Space Administration.

This publication also makes use of data from SDSS-III. Funding for SDSS-III has been provided by the Alfred P. Sloan Foundation, the Participating Institutions, the National Science Foundation, and the U.S. Department of Energy Office of Science. The SDSS-III web site is \url{http://www.sdss3.org/}.

SDSS-III is managed by the Astrophysical Research Consortium for the Participating Institutions of the SDSS-III Collaboration including the University of Arizona, the Brazilian Participation Group, Brookhaven National Laboratory, Carnegie Mellon University, University of Florida, the French Participation Group, the German Participation Group, Harvard University, the Instituto de Astrofisica de Canarias, the Michigan State/Notre Dame/JINA Participation Group, Johns Hopkins University, Lawrence Berkeley National Laboratory, Max Planck Institute for Astrophysics, Max Planck Institute for Extraterrestrial Physics, New Mexico State University, New York University, Ohio State University, Pennsylvania State University, University of Portsmouth, Princeton University, the Spanish Participation Group, University of Tokyo, University of Utah, Vanderbilt University, University of Virginia, University of Washington, and Yale University.

}


\end{document}